\documentclass{article}

\usepackage{arxiv}

\usepackage[utf8]{inputenc} 
\usepackage[T1]{fontenc}    
\usepackage{hyperref}       
\usepackage{url}            
\usepackage{booktabs}       
\usepackage{amsfonts}       
\usepackage{nicefrac}       
\usepackage{microtype}      
\usepackage{lipsum}
\usepackage{graphicx}
\graphicspath{ {./images/} }
\usepackage{array}
\usepackage{amsmath}
\usepackage{float}
\usepackage{subcaption}
\usepackage[font=small,labelfont=bf]{caption}
\usepackage{multirow}
\usepackage{forest}
\forestset{qtree/.style={for tree={parent anchor=south, 
           child anchor=north,align=center,inner sep=0pt}}}
\newcommand{\indep}{\raisebox{0.05em}{\rotatebox[origin=c]{90}{$\models$}}}

\title{Causal Inference in Educational Systems: A Graphical Modeling Approach}

\author{
  Manie Tadayon\\
  Department of Electrical and Computer Engineering\\
  University of California, Los Angeles\\
  Los Angeles, CA  \\
  \texttt{manitadayon@ucla.edu} \\
   \And
  Greg Pottie \\
  Department of Electrical and Computer Engineering\\
  University of California, Los Angeles\\
  Los Angeles, CA \\
  \texttt{pottie@ee.ucla.edu} \\
}
\begin{document}
\maketitle
\begin{abstract}
Educational systems have traditionally been evaluated using cross-sectional
studies, namely, examining a pretest, posttest, and single intervention. Although this is a popular approach, it does not model valuable information such as confounding variables, feedback to students, and other real-world deviations of studies from ideal conditions. Moreover, learning inherently is a sequential process and should involve a sequence of interventions. In this paper, we propose various experimental and quasi-experimental designs for educational systems and quantify them using the graphical model and directed acyclic graph (DAG) language. We discuss the applications and limitations of each method in education. Furthermore, we propose to model the education system as time-varying treatments, confounders, and time-varying treatments-confounders feedback. We show that if we control for a sufficient set of confounders and use appropriate inference techniques such as the inverse probability of treatment weighting (IPTW) or g-formula, we can close the backdoor paths and derive the unbiased causal estimate of joint interventions on the outcome. Finally, we compare the g-formula and IPTW performance and discuss the pros and cons of using each method. 
\end{abstract}


\section{Introduction}
\label{sec:headings}
Educational systems have traditionally been evaluated using cross-sectional
studies, namely, examining a pretest and posttest, and single intervention. Although this is a popular approach, it does not model valuable information such as confounding variables, feedback to students, and other real world deviations of studies from ideal conditions. Moreover, learning inherently is a sequential process and should involve a sequence of interventions. Nowadays, a large volume of educational data is available, allowing researchers to develop more intelligent and more complex modeling and inference algorithms.\\
Considerable research such as \cite{anderson1985intelligent}, \cite{nwana1990intelligent}, and \cite{conati2009intelligent} has been done in designing intelligent tutoring systems (ITS) in education. ITS is computer software that dynamically evaluates student knowledge and provides feedback. Some work such as \cite{tadayon2020predicting} and \cite{carmona2008designing} has been done in modeling student knowledge using hidden Markov model (HMM) and dynamic Bayesian networks (DBN). In \cite{tadayon2020predicting}, the authors used the HMM to predict student performance in an educational video game. They showed that student trajectories throughout the game can be strongly predictive of posttest scores. Some work such as \cite{yudelson2013individualized}, \cite{corbett1994knowledge}, and \cite{piech2015deep} has been done in the knowledge tracing, Bayesian knowledge tracing, and deep knowledge tracing. In \cite{piech2015deep} the authors used a recurrent neural network (RNN) for deep learning knowledge tracing to model student learning. Although these frameworks are popular in education, they will not model time-varying confounders and factors that can influence student learning. \\
Potential outcome (PO) is a well-known causal inference framework in literature. Neyman first proposed PO for randomized experiments, and later on, it was extended to observational and non-randomized experiments by Donald Rubin. 
Causal DAG and structural causal model (SCM) developed by Judea Pearl \cite{pearl2009causality} is another popular causal inference framework that uses directed graphs to model real world phenomena. Directed graphs have been used in many domains \cite{bessani2020multiple}, \cite{zuo2012stock}, \cite{queen2009intervention}, and \cite{tadayon2020tsbngen}. In \cite{tadayon2020tsbngen}, the authors used dynamic Bayesian networks to generate synthetic time series data.\\
In this paper, first, we use causal DAG and graphical modeling frameworks to quantify the experimental and quasi-experimental studies. We study some well-known experimental and quasi-experimental designs in education and discuss their limitations in terms of threats to internal and external validities mentioned by \cite{cook2002experimental}.
Second, we propose more intelligent experimental designs by collecting richer data from students by including questions on potential confounders in the diagnostic test and instructing the TAs and instructors to inquire during office hours about potential confounders. Third, we propose to model the education system as time-varying treatments, confounders, and time-varying treatments-confounders feedback. We show that if we control for a sufficient set of confounders and use appropriate techniques such as the marginal structural model along with IPTW \cite{robins2000marginal} or g-formula, which is the extension of Pearl's backdoor criterion, we can close a sufficient set of backdoor paths and derive the unbiased causal estimate of joint interventions on the outcome. \\
The rest of this paper is organized as follows. Section 2
reviews background materials used in other sections. Section 3 describes the experimental designs in education. Section 4 describes the non-randomized designs with a single intervention in education. Section 5 describes the non-randomized studies with multiple interventions where we propose time-varying treatment-confounder feedback and many other models in education. Section 6 presents the conclusion.

\section{Background}
\label{Background}
\begin{center}
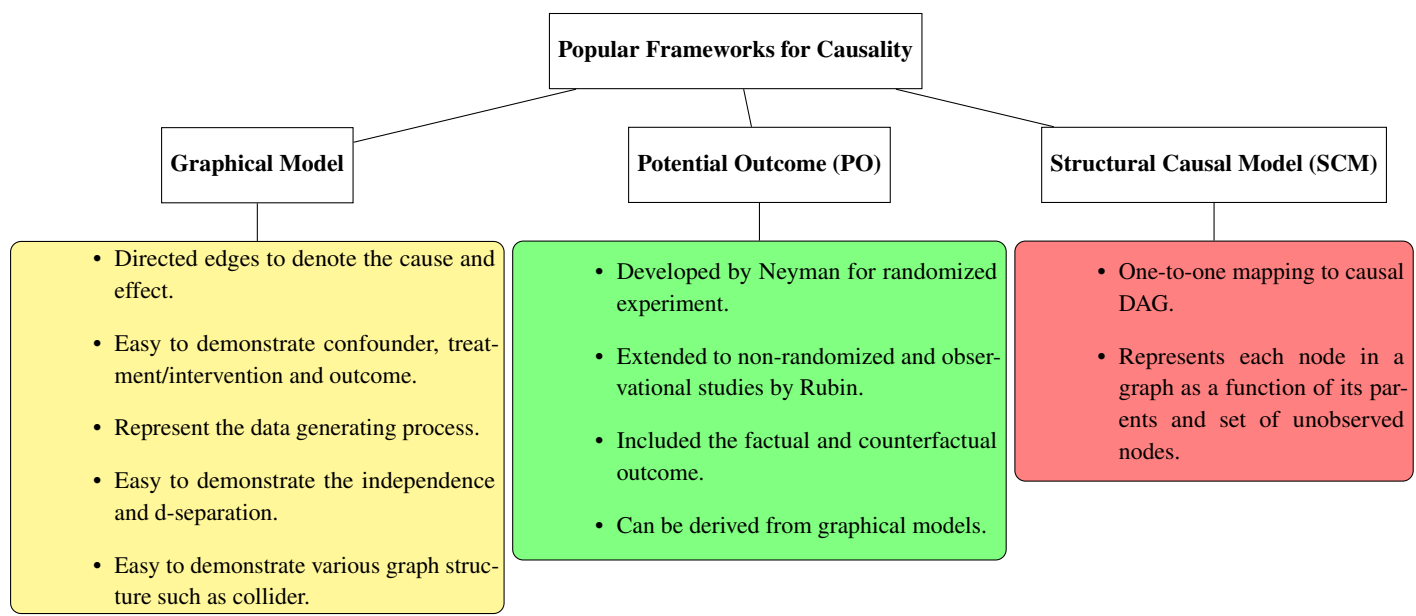
\begin{figure}[tbh]
\begin{forest}
for tree={%
    l sep=0.5cm,
    s sep=0.1cm,
    minimum height=1cm,
    minimum width=1.5cm,
    font=\footnotesize,
    draw 
    }
 [\textbf{Popular Frameworks for Causality}
 [\textbf{Graphical Model} [\parbox{20em} {\begin{itemize}
     \item Directed edges to denote the cause and effect.
     \item Easy to demonstrate confounder, treatment/intervention and outcome.
     \item Represent the data generating process.
     \item Easy to demonstrate the independence and d-separation.
     \item Easy to demonstrate various graph structure such as collider.
 \end{itemize}},fill=yellow!50, rounded corners]]
 [\textbf{Potential Outcome {(PO)}} [\parbox{20em} {\begin{itemize}
     \item Developed by Neyman for randomized experiment.
     \item Extended to non-randomized and observational studies by Rubin.
     \item Included the factual and counterfactual outcome.
     \item Can be derived from graphical models.
 \end{itemize}},fill=green!50, rounded corners]]
 [\textbf{Structural Causal Model {(SCM)}}[\parbox{16em} {\begin{itemize}
     \item One-to-one mapping to causal DAG.
     \item Represents each node in a graph as a function of its parents and set of unobserved nodes.
 \end{itemize}},fill=red!50, rounded corners]]
 ]
\end{forest}
 \caption{Well-known Frameworks for Causality}\label{Causality_Framework1}
\end{figure}
\end{center}

\begin{figure}[tbh]
    \centering
    \includegraphics[width = 0.70\textwidth]{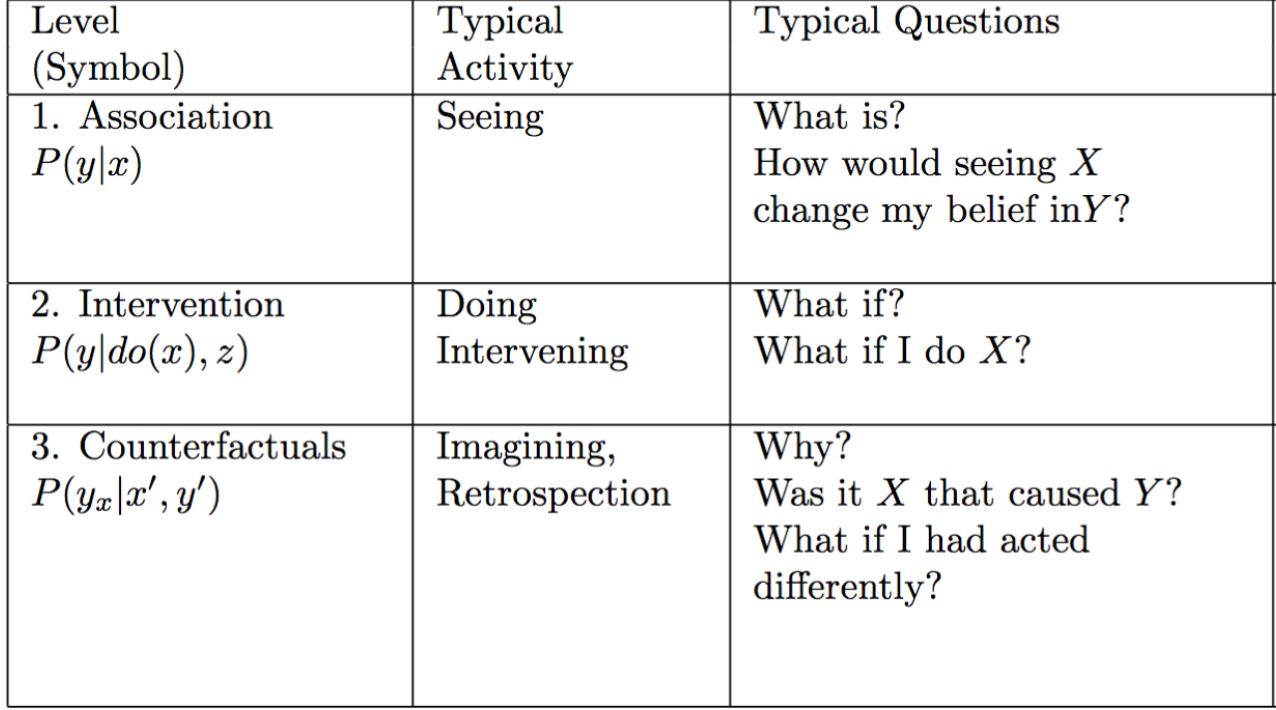}
    \caption{Pearl Ladder of Causation}
    \label{Pearl Ladder}
\end{figure}
\vspace{3cm}

\subsection{Pearl Ladder of Causation}
The Following are the rungs in Pearl's hierarchical ladder of causation: 
\begin{itemize}
    \item \textbf{Association}: This is the first layer in the ladder of causation corresponding to the first row of figure \ref{Pearl Ladder}. This is where  most of the current statistical literature lies. An example of association or correlation is the conditional probability of $P(y|x)$, which means the probability of obtaining $y$ given we observe or see $x$.
    \item \textbf{Intervention}: This is the second layer in ladder of causation corresponding to the second row of figure \ref{Pearl Ladder}. This is the layer responsible for the do calculus. An example of this is $P(y|do(x))$, which means the probability of obtaining $y$ given we are forcing or intervening $X = x$.\\
    \textbf{Note:} $P(y|do(x))$ is commonly written as $P(y_x)$ (Potential outcome notation, explained later in this section).
    \item \textbf{Counterfactual}: This is the third layer in the ladder of causation corresponding to the third row of figure \ref{Pearl Ladder}. The counterfactual output is always unobserved. For example, consider $x = 1$ denotes receiving the treatment and $x = 0$ denotes otherwise. Then $P(y_{x = 1}| x = 0)$ is a counterfactual outcome since we are asking about the probability of outcome among the people who did not receive the treatment had we forced them to receive the treatment.
\end{itemize} 
\subsection{Potential Outcome (PO)}
$Y^x$ or $Y_x$ is the outcome that would be observed if the treatment was set to X=x; in other words, it is the outcome under different possible treatment options. It consists of observed or factual and unobserved or counterfactual segments. \\
Assume the case of binary treatments, where $X = 1$ denotes receiving the treatment and $X = 0$ denotes otherwise. $C = c$ denotes observing participants among those that are in group $c$. For example, in education, $c$  can represent students who are in a specific class, have a certain GPA, etc. Then 
based on the definition of the PO and the counterfactual in the last section, we can define the following equations.
\begin{align}
    &\textrm{Average Treatment Effect} =  ATE  = E[y_{x=1}] - E[y_{x=0}] \\
    &\textrm{Average Treatment Effect Among Treated} =  ATT = E[y_{x=1}|X = 1] - E[y_{x=0}|X = 1] \\ 
    &\textrm{Average Treatment Effect Among Untreated} =  ATU = E[y_{x=1}|X = 0] - E[y_{x=0}|X = 0] \\ 
      &\textrm{Conditional Treatment Effect} = CATE = E[y_{x=1}|C = c] - E[y_{x=0}|C = c] 
\end{align}
\subsection{Graphical Model (Causal DAG) }
\begin{figure}[tbh]
    \centering
    \includegraphics[scale=0.8]{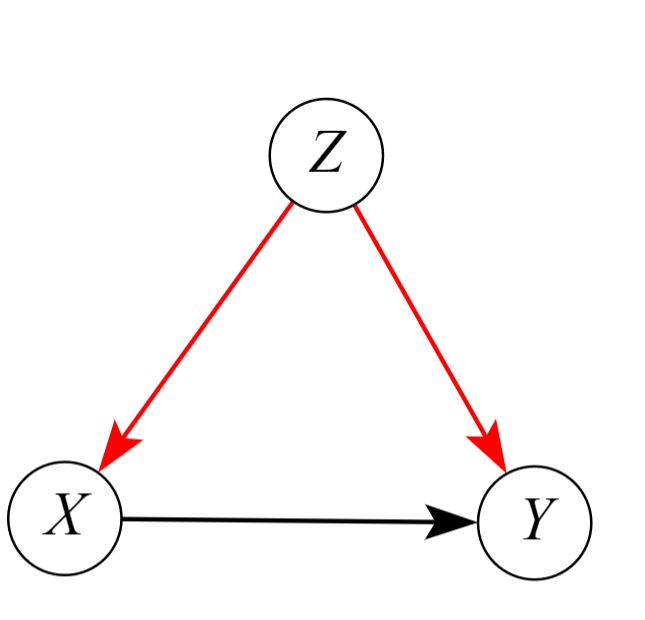}
    \caption{Graphical Representation of Treatment, Outcome and Confounder}
    \label{backdoor1}
\end{figure}

 In this framework, the nodes and variables are represented as vertices of the graph, and directed edges denote causal relationships between them. Besides having directed edges, there is no cycle or loop in a causal DAG. The benefit of using a DAG is to define a pictorial view of the problem, to represent the data generating process, and to demonstrate independence and d-separation clearly. For example, in figure \ref{backdoor1}, $X$ is the treatment, $Y$ is the outcome, and $Z$ is the confounder (cause of both $X$ and $Y$ and is placed on the backdoor path between $X$ to $Y$). In causal inference, we are usually interested in finding the average causal effect of treatment on outcome. If $Z$ were not present, then the problem would be reduced to correlational studies such that causation and correlation would be the same. \\
In figure \ref{backdoor1}, the red path $X \leftarrow Z \rightarrow Y $ is called the backdoor path, which is a non-causal path pointing to a treatment $X$ and outcome $Y$. Our goal in causal inference is to block backdoor paths since they are non-causal paths and introduce bias in our analysis. To account for this backdoor path and determine unbiased causal estimate of $X$ on $Y$ we use the Pearl backdoor criterion as follows:\\
\textbf{Backdoor Adjustment}: If Z is a backdoor variable relative to $X$ and $Y$ then the causal effect of $X$ on $Y$ is calculated by:
\begin{align}
    P(Y|do(X)) = \sum_Z P(Y|X,Z) P(Z) 
    \label{backdoor2}
\end{align}
\subsection{ Structures in DAG}
\begin{figure}[tbh]
\centering
\begin{subfigure}{.33\textwidth}
  \centering
  \includegraphics[width=0.8\linewidth]{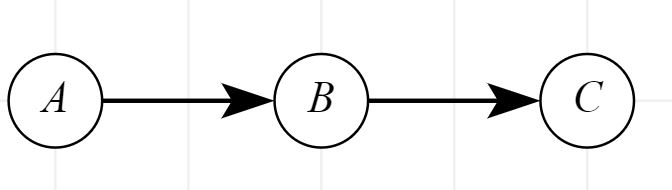}
  \subcaption{$A \indep C|B$}
\end{subfigure}%
\begin{subfigure}{.33\textwidth}
  \centering
  \includegraphics[width=0.7\linewidth]{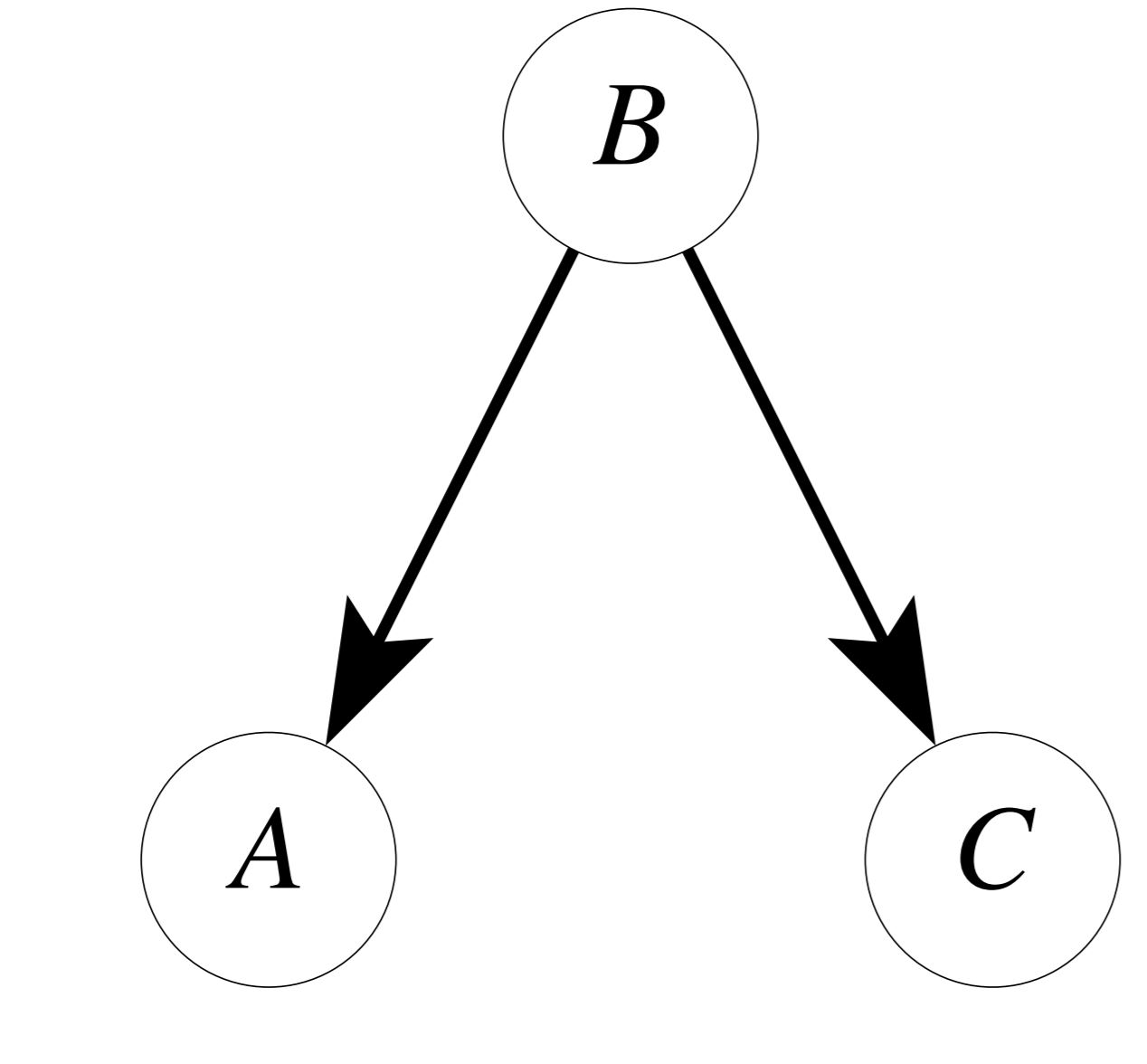}
   \subcaption{$A \indep C|B$}
\end{subfigure}
\begin{subfigure}{.33\textwidth}
  \centering
  \includegraphics[width=0.7\linewidth]{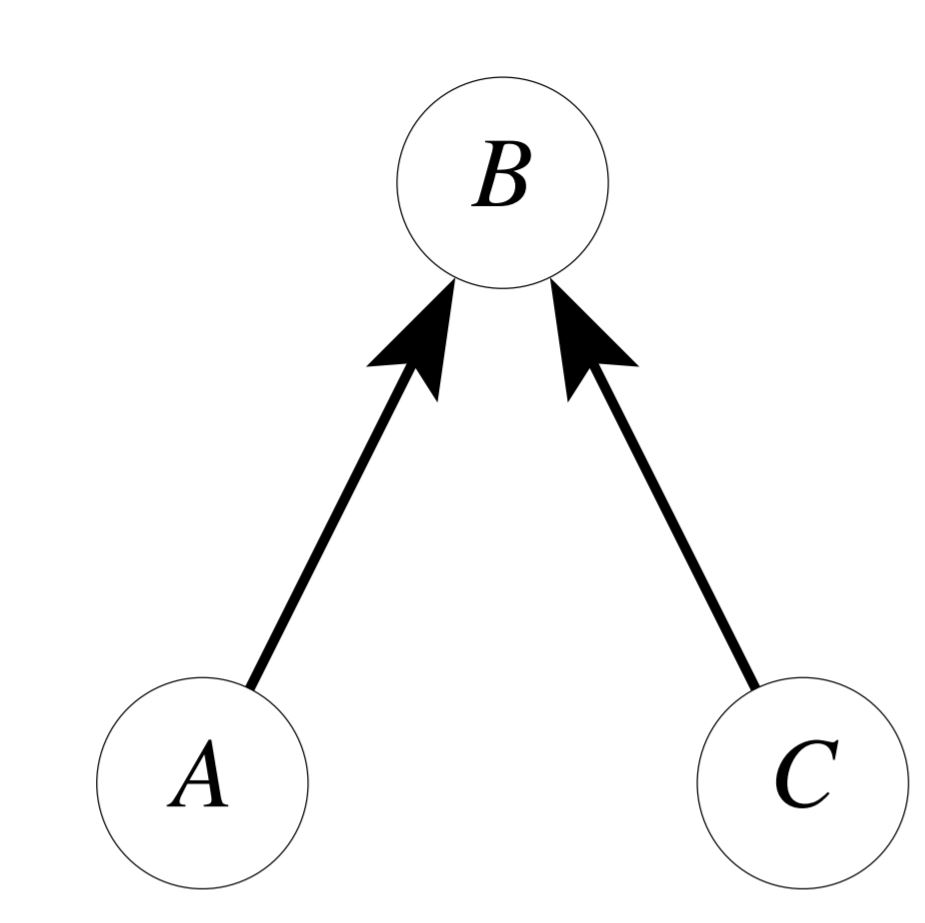}
   \subcaption{$A  \indep C$}
\end{subfigure}
\caption{Various structures on a DAG}
\label{Structure1}
\end{figure}

Figure \ref{Structure1} shows various independence structures on DAGs. Figures \ref{Structure1}a and \ref{Structure1}b are structures in which $A$ and $C$ are d-separated conditioned on B. However \ref{Structure1}c denotes a collider structure, where $A$ and $C$ are independent but they will be dependent conditioned on $B$. Conditioning on $B$ results in what is known as a collider stratification bias.
\subsection{Pre-Intervention and Post-Intervention DAG}
\begin{figure}[tbh]
    \centering
    \includegraphics{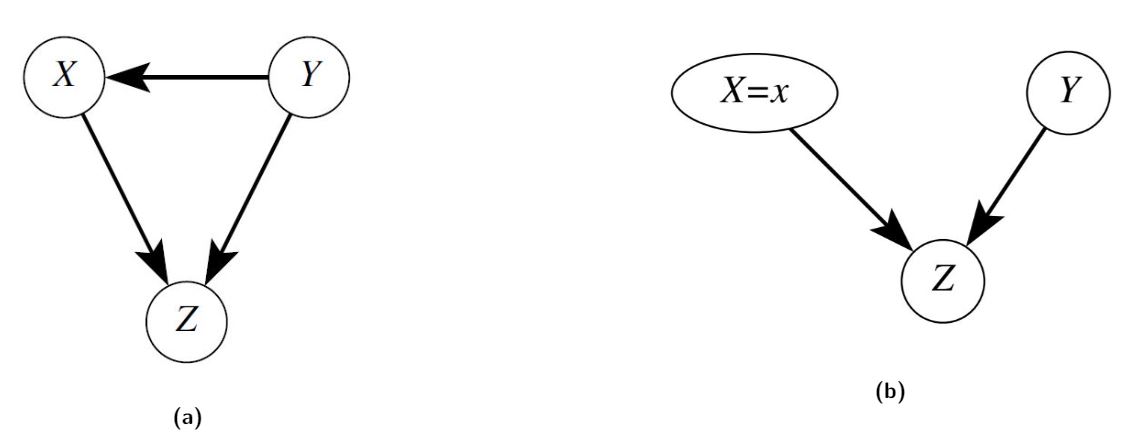}
    \caption{\textbf{a)} The Pre-Intervention DAG \quad \textbf{b)} The Post-Intervention DAG }
    \label{pre-post_DAG}
\end{figure}

The pre-intervention DAG is a data generating process, or a DAG designed by an expert or given by the nature. The post-intervention DAG is a pre-intervention DAG after performing the intervention. This is done by removing the parents of the nodes we are doing an intervention on. For example figure \ref{pre-post_DAG}a shows a pre-intervention DAG and \ref{pre-post_DAG}b shows the post-intervention DAG after performing the intervention on $X$. 
\subsection{Structural Causal Model (SCM)}
The SCM is a system of structural equations relating each node in a DAG to their parents and the set of unobserved causes. It has a one-to-one mapping to graphical models. The following are the components of the SCM model:
\begin{itemize}
    \item Variables/Nodes.
    \item Unobserved cause of nodes.
    \item Functions mapping each node to their parents.
\end{itemize}
For example, in figure \ref{pre-post_DAG}, the following equations demonstrate the SCM for the pre (the left equation) and post (the right equation) intervention DAGs.
    \begin{align}
     Y &= f_{Y}(U_{Y})       &  Y & = f_{Y}(U_{Y}) \\
     X &= f_{X}(Y,U_{X})     &  X & = x \\
     Z &= f_{Z}(X,Y,U_{Z})   &  Z &= f_{Z}(x,Y,U_{Z}) \\ \quad \quad &U_X \indep U_Y \indep U_Z
    \end{align}

\subsection{Identifiability Condition}
\begin{itemize}
    \item Well-Specified Variables.
    \begin{itemize}
        \item All variables should be properly defined. 
        \item The treatment, outcome and measured confounders should be correctly specified.
    \end{itemize}
    \item Positivity.
    \begin{itemize}
        \item No deterministic treatment within levels of the confounder. For example, in figure \ref{backdoor1}, for positivity to hold we need:
        \begin{align}
            P(X = x| Z =z) > 0 \quad \quad \textrm{If} \quad P(Z = z)>0
            \label{positivity}
        \end{align}
        If the equation \ref{positivity} is violated then $E(Y|X,Z)$ is not defined. This is because:
        \begin{align}
            E(Y|X,Z) = \sum_y y \hspace{0.1cm}p(y|X,Z) = \sum_y y \dfrac{p(y,x,z)}{p(x,z)} = \sum_y y \dfrac{p(y,x,z)}{p(x|z)p(z)}
            \label{positivity2}
        \end{align}
        since the numerator of equation \ref{positivity2} will be zero.
    \end{itemize}
    \item Consistency.
    \begin{itemize}
        \item For those who receive $X =x$ then $Y_x = y$.
        \item For those who receive $X =x^{\star}$ then $Y_{x^{\star}} = y$.\\
        For example using consistency, $E[Y_x]$ can be written as follows:
        \begin{align}
            &E(Y_x) = \sum_x E(Y_x|X = x) p(x) = E(Y_x|X = x) p(x) + E(Y_x|X = x^{\star}) p(x^{\star}) \\
             & =  E(Y|X = x) p(x) + E(Y_x|X = x^{\star}) p(x^{\star})
        \end{align}
    \end{itemize}
    \item Conditional Exchangeability.
    \begin{itemize}
    \item Conditional exchangeability states that the potential outcome is independent of the treatment given the sufficient set of confounders. Mathematically this is represented as follows:
    \begin{align}
        Y_X \indep X | Z
        \label{cond_exchan}
    \end{align}
    Conditional exchangeability states if you account for sufficient confounders then treatment and control groups are the same.
    \item Z in equation \ref{cond_exchan} is sufficient for controlling any non-causal path.
    \item It states that the treated and untreated groups are similar conditioned on the set Z.
    \end{itemize}
    \item Defining Sources of Bias.
    \begin{itemize}
        \item No measurement error.
        \item No selection bias (loss-to-follow-up, etc).
    \end{itemize}
    \item Well-specified Model.\\
    Define a valid statistical model for inference and testing. Make sure the statistical model can match the problem requirements in terms of amount of data, number of parameters, nature of data (whether data is cross-sectional or longitudinal), etc.
\end{itemize}

\subsection{Internal and External Validities}
External validity is the ability to generalize the findings to other populations and settings. Lack of external validity suggests that the outcome cannot be related to other people or contexts outside the current study. For example, consider studies where the participants are all male or students who all have perfect GPA, then the findings cannot be generalized to the entire population.\\
Table \ref{External_table} summarizes some well-known threats to external validities.
\begin{table}[tbh]
\centering
\caption{Threats to External Validity}
\begin{tabular}{ |c|c| }
\hline
 \textbf{Threats} & \textbf{Definition} \\ 
 \hline
 \textbf{Participation Characteristics} &  Participants should be a good representative (sample) of entire population. \\  
 \hline
  \textbf{Experiment Setting} & How similar is the experiment setting to real-world. \\
 \hline
 \textbf{Timing} & Study over the past might not be the same as now. \\
 \hline
\end{tabular}
\label{External_table}
\end{table}

Internal validity is defined as the degree to which the causal relationship between dependent and independent variables can be established. Internal validity is trying to answer the following questions: Is the independent variable the only cause of the change in the dependent variable? Have we considered all the confounders in the study? Table \ref{Internal_table} indicates some well-known threats to internal validity according to \cite{cook2002experimental}.
\begin{table}[h]
\centering
\caption{Threats to Internal Validity}
\begin{tabular}{ |c|c| }
\hline
 \textbf{Threats} & \textbf{Definition} \\ 
 \hline
 \textbf{Ambiguous Temporal Precedence} &  Lack of clarity between which variable occurs first. \\  
 \hline
  \textbf{Regression} & Problem of regression to the mean.  \\
 \hline
 \textbf{Confounding Variables} & Something other than treatment is causing the change in outcome. \\
 \hline
  \textbf{History} & External events that affect how participants responds. \\
 \hline
  \textbf{Maturation} & Natural process in which participants change during the course of study. \\
 \hline
  \textbf{Instrumentation} & An instrument or a measure in a study change over the course of study . \\
 \hline
  \textbf{Testing/Practice Effects} & Repetition of test to enhance the performance. \\
 \hline
  \textbf{Attrition/Mortality} & Participants dropping out during the course of study. \\
 \hline
\end{tabular}
\label{Internal_table}
\end{table}
\section{Experimental Designs in Education}
\begin{figure}[h]
\centering
\begin{subfigure}{.5\textwidth}
  \centering
  \includegraphics[width=.9\linewidth]{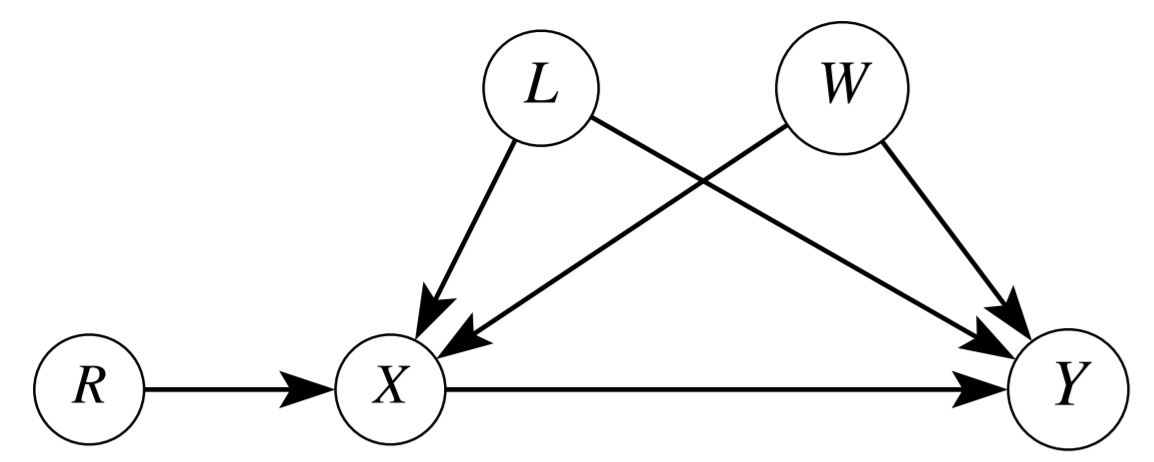}
  \subcaption{}
\end{subfigure}%
\begin{subfigure}{.5\textwidth}
  \centering
  \includegraphics[width=.7\linewidth]{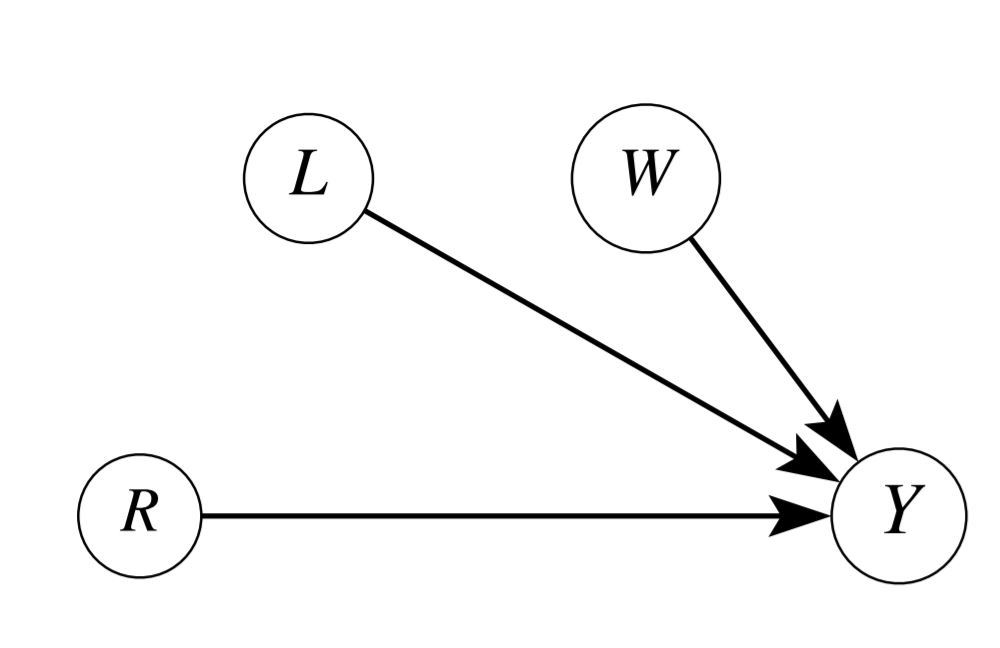}
   \subcaption{}
\end{subfigure}
\caption{a) DAG for a RCT with non-compliance. b) DAG for RCT with full compliance}
\label{RCT1}
\end{figure}

Experimental designs are designs where the researcher has complete control over the study. Their main characteristic is randomization, where the participants are assigned to control and treatment groups randomly. Perhaps the most well-known type of experimental design is the randomized controlled trial (RCT). It assigns the participants to the control and the treatment groups randomly. Hence, it will remove the confounding variables from the study. The ideal RCT (full compliance) is considered the gold standard in medicine; however, in practice, participants might leave the study or do not comply with the treatment, which creates a bias in the study. Figure \ref{RCT1}a demonstrates a candidate DAG for the RCT Under the non-compliance where $R$ is the instrument and denotes the randomization to receive the treatment, $X$ is whether you actually receive the treatment, $L$ and $W$ are the confounders (They may be observed or unobserved), and Y is the outcome. For example, in education, $X$ can be either solving some problems, reading sections of a book, or getting a referral to an expert; $L$ and $W$ are either the observed or unobserved confounders such as family issues, socioeconomic status, prior knowledge and stress, and psychological problems. $Y$ denotes the outcome of interest, for example, the exam score. In figure \ref{RCT1}a $X$ is a collider, hence conditioning on $X$ will open the non-causal path $R\rightarrow X\leftarrow L\rightarrow Y$ and $R\rightarrow X\leftarrow W\rightarrow Y$.\\
Figure \ref{RCT1}b shows the same DAG but an ideal RCT, where the effect of confounding variables is removed. In figure \ref{RCT1}b, $R$ and $X$ are the same, since if you are randomized to receive the treatment, then you will receive the treatment (Full-compliance)\\
However, even if the RCT is done with full compliance, it still suffers from the following limitations:
\begin{itemize}
    \item It is expensive. In education usually funding is limited, which makes the RCT not practical as a tool for widespread use.
    \item It requires careful planning and correct strategy to keep participants in the study and perform the correct randomization.
    \item It is unfeasible and unethical in many situations. For example, you may not be able to force students talk to the TA or visit a counselor.
\end{itemize}

\begin{figure}
        \centering
        \includegraphics[width=0.7 \textwidth]{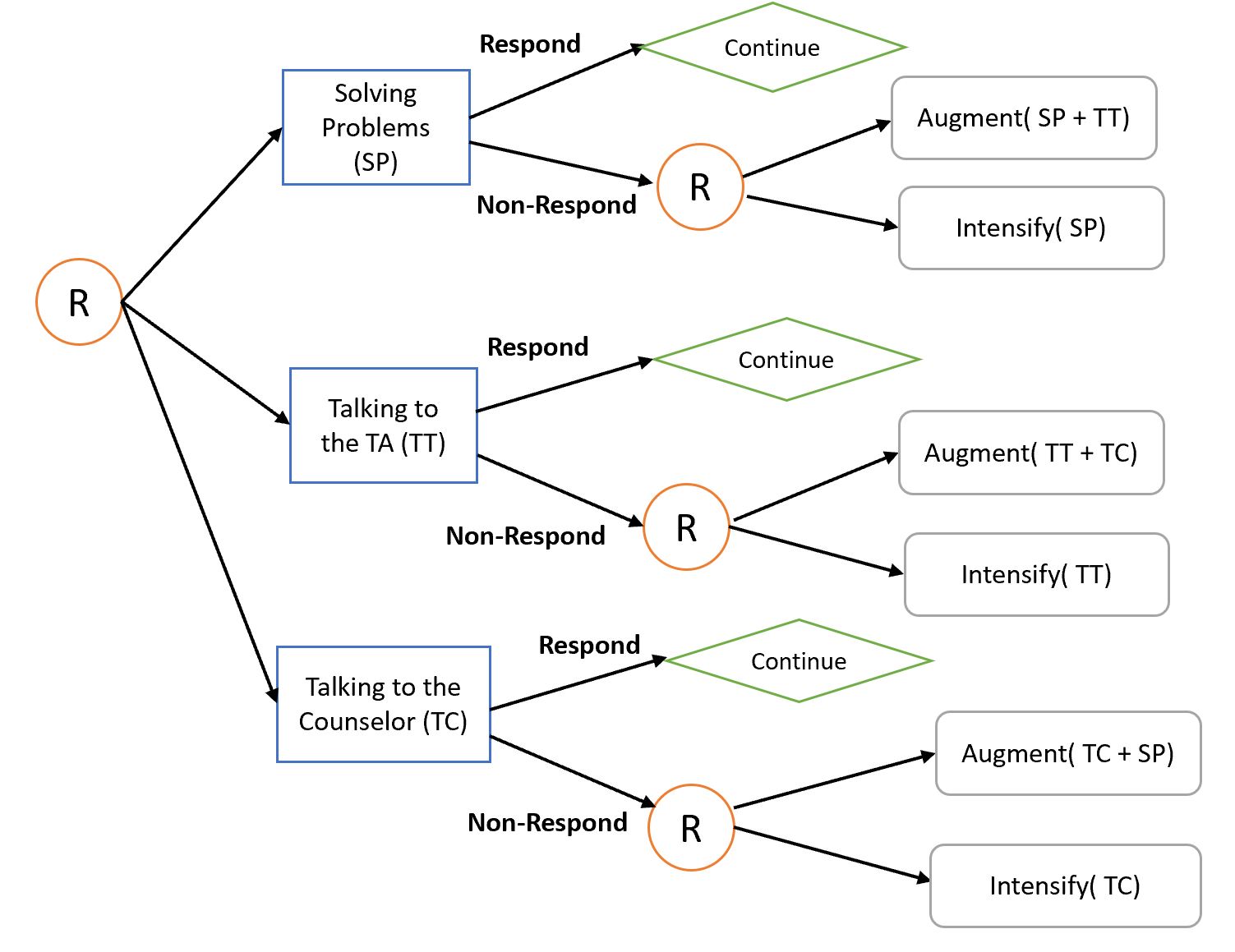}
        \hfill
      \caption{ SMART data collection design in education }
        \label{SMART}
    \end{figure}

Sequentially Multiple Assignment Randomized Trial (SMART) \cite{collins2007multiphase} is another type of experimental design that consists of randomization at each stage. Figure \ref{SMART} shows a proposed SMART design in education, where initially students are randomly assigned to three groups based on interventions they receive:1- Students who are recommended to solve some problems (SP), 2- Students are recommended to talk to the TA (TT). 3 - Students who are recommended to talk to a counselor (TC). Students then continue with the intervention for some duration, such as five weeks, and are assessed after that. Those who respond to treatment, meaning that treatment is effective for them, continue with this treatment, and those who are non-respondent will be randomized again to students who receive the augmented treatments, namely combining the previous treatment with the new one and students who receive the previous treatment with the higher intensity. This process may continue until a certain mastery level for all students is achieved or until the duration of study is over. Figure \ref{SMART} shows a two-stage SMART design with three treatment options at the first stage.

Since SMART is a multi-stage randomization design, then similar to the RCT, it is costly and requires a careful planning. Therefore it may not be a practical option for educational design. In the next sections, we will discuss non-randomized (quasi-experimental) designs in education.
\section{Non-randomized Designs with a Single Intervention}
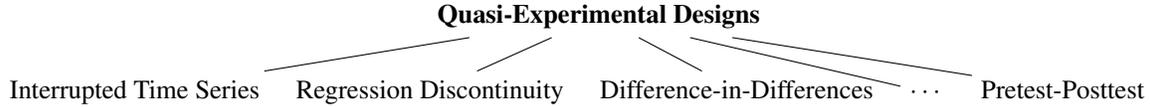
\begin{figure}[tbh]
\centering
\begin{forest}
 [\textbf{Quasi-Experimental Designs}[Interrupted Time Series][Regression Discontinuity][Difference-in-Differences] [$\cdots$] [Pretest-Posttest]]
 ]
\end{forest}
\caption{Some Well-Known Quasi-Experimental Designs}
\label{QED1}
\end{figure}

Non-randomized or quasi-experimental designs are designs used when the randomized experiment is not possible due to ethical, timing, or cost restrictions. They are sometimes also referred to as observational studies. \cite{rosenbaum2005observational}.\\
Figure \ref{QED1} represents some well-known quasi-experimental designs in literature:\\
In \cite{cook2002experimental}, the authors introduced a four levels of hierarchy for quasi-experimental designs demonstrated in table \ref{QED2_cook}. In Table \ref{QED2_cook}, D is the best design, followed by C, B and A. Next, we discuss the use of designs in table \ref{QED2_cook} in educations.
\begin{table}[tbh]
\centering
\caption{Hierarchy of Quasi-Experimental Design}
\begin{tabular}{ |l| }
\hline
 \textbf{Type of Experiment} \\ 
 \hline
 A. Design with no control groups. \\  
 \hline
 B. Design with control groups but no pretest. \\
 \hline
 C. Design with control groups and pretest \\
 \hline
 D. Design for time series data.\\
 \hline
\end{tabular}
\label{QED2_cook}
\end{table}\\
\textbf{Note}: The common notation is to use X for intervention, O for measurement (observation), and the NR for non-randomized assignment.The position from left to right determines the temporal order.\\ \\
\subsection{\textbf{Designs with no control groups}}
1. \textbf{The One group posttest design:} \par
     \hspace{3cm}    X      \quad \quad $O_1$
\par
\begin{figure}[tbh]
    \centering
    \includegraphics[width=0.5\textwidth]{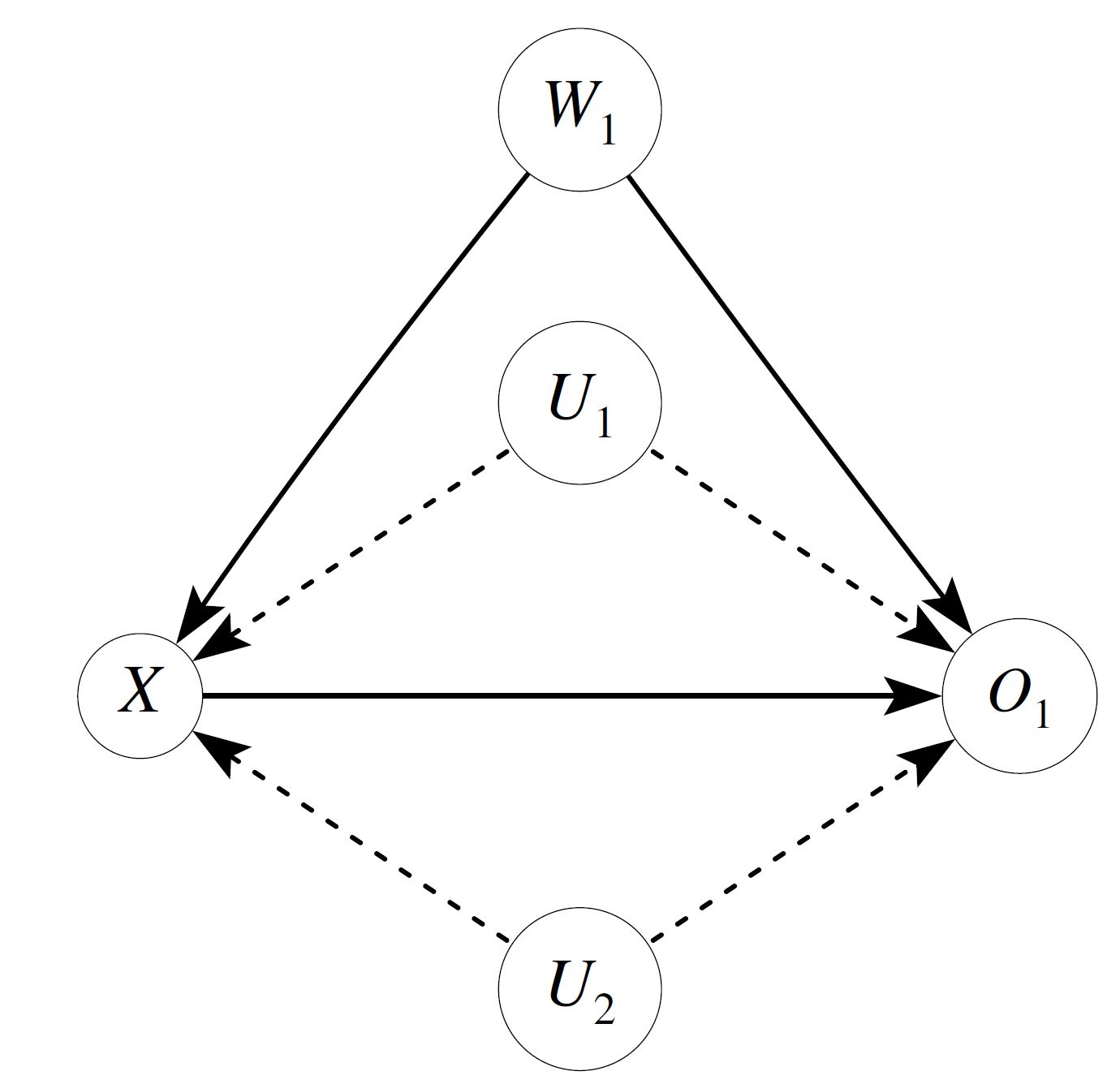}
    \caption{DAGs for a single posttest design}
    \label{posttest}
\end{figure}

  Figure \ref{posttest} shows an example of a DAG for this scenario. X is a treatment or an intervention, $O_1$ is an outcome or the posttest, $W_1$ is observed, and $U_1$ and $U_2$ are unobserved confounders. For example, suppose a researcher is interested in finding the causal effect of doing some problem sets in a textbook on student learning by taking a final exam or posttest immediately after the study ends. Then  $X$ is doing the problem sets, and the $O_1$ is the final exam. This design is subject to many threats to internal validities such as history since other events besides the treatment might cause the change in exam score. Confounding variables such as prior knowledge of the topic, intelligence, family, and psychological problems (unobserved confounder) might cause a change in outcome. Lack of pretest does not let us query students' prior knowledge and the absence of a control group, makes it difficult to determine if treatment is the only cause of change in the outcome. To better understand the problem with this design, consider the DAG in figure \ref{posttest}; ideally, we would like to find the direct causal effect of $X$ on $O_1$; however, the X $\leftarrow$ $\huge W_1$ $\rightarrow$ $\huge O_1$ is a non-causal backdoor path, which should be closed. But, even blocking the backdoor path through $W_1$ still is not sufficient due to the backdoor path via unobserved confounders.\\
To better understand the problem with this design consider the following simulation. The DAG for the data generating process is the same as figure \ref{posttest}. The treatment and all the confounders (observed and unobserved) are Bernoulli random variables and the outcome $O_1$ is normally distributed with mean to be the linear combination of $X$, $U_1$, $U_2$ and $W_1$ and the standard deviation is 0.6.\\
\textbf{Simulation Setup} \\
\begin{align*}
    N=& 500 \\
    P(W_1 = 1) =& 0.8 \\
    P(U_1 = 1) =& 0.4 \\
    P(U_2 = 1) =& 0.6 \\
    P(X =1) =& 0.30 + 0.2U_1 + 0.5U_2 - 0.3W_1 \\
    \mu=&  5 + 7X + 4W_1 - 2U_1 - 2U_2  \\
    \sigma =& 0.6 \quad \quad O_1 \sim N(\mu,\sigma^2)\\
\end{align*}
\begin{table}[tbh]
\centering
\caption{Simulation Result for a design with a single intervention}
    \begin{tabular}{|c|c|c|}
    \hline
    \cline{1-3}
    \multicolumn{3}{|c|}{\textbf{Control for No Confounders}}\\
    \hline
         {Coefficient} & {True} & {Estimate}  \\
         \hline
            Intercept & 5.0 & 7.13 \\ 
            \hline
            $X$ &  7.0  &  5.04\\
            \hline
      \multicolumn{3}{|c|}{\textbf{Control for $W_1$}}\\ 
      \hline
            Intercept & 5.0 & 3.88 \\ 
            \hline
            $X$ &  7.0  &  5.58\\
            \hline
     \multicolumn{3}{|c|}{\textbf{Control for $W_1$ and $U_1$}}\\ 
      \hline
            Intercept & 5.0 & 4.27 \\ 
            \hline
            $X$ &  7.0   &  6.55\\
            \hline
      \multicolumn{3}{|c|}{\textbf{Control for $W_1$, $U_1$ and $U_2$}}\\ 
      \hline
            Intercept & 5.0 & 4.99 \\ 
            \hline
            $X$ &  7.0   &  7.03\\
            \hline
    \end{tabular}
\label{single_intervention_table} 
\end{table}
Table \ref{single_intervention_table} presents simulation results for this design. According to this table, the causal effect of $X$ on $O_1$ will be biased if we do not control for unobserved confounders.\\ 
\par
2. {\textbf{The One-Group Pretest-Posttest Design:} \par
     \hspace{3cm} $O_1$ \quad \quad   X      \quad \quad $O_2$
\par Figure \ref{pre-post2}a is a candidate DAG for this case, where the outcome $O_2$ is the posttest score (It can be a continuous or discrete quantity). $U_1$ and $U_2$ are unobserved confounders and can refer to psychological and family problems. $W_1$ represents any observed confounder. This is similar to the DAG in figure \ref{posttest}. This is because, in causal inference, we study the causal effect of intervention/treatment on the outcome and the pretest-posttest design has similar characteristics as one-group posttest design after the intervention. Including a pretest, however, can provide some information about the student's prior knowledge, which in turn helps us to control for some confounders compared to the first design. However, repeating the same problems in the consecutive exams results in testing/practice effects that threatens internal validity. This design is also susceptible to other threads to internal validities such as maturation and history. This is because participants may drop out during the study (for example, after taking the pretest). This is sometimes referred to as a loss-to-follow-up, which causes the selection bias. This design also suffers from regression to the mean since if there is only one observation before the intervention (the pretest), then the score in the pretest can be due to chance. Therefore, the pretest would be a very weak counterfactual had the treatment never happened. Figure \ref{pre-post2}b is another candidate DAG, which includes the pretest $O_1$ and $L_1$ as a covariate causing students to drop from the study. $C_1$ is a binary variable that represents the censoring where $C_1 = 0$ students are not censored and continue the study.\\
\begin{figure}[h]
\centering
\begin{subfigure}{.4\textwidth}
  \centering
  \includegraphics[width=.9\linewidth]{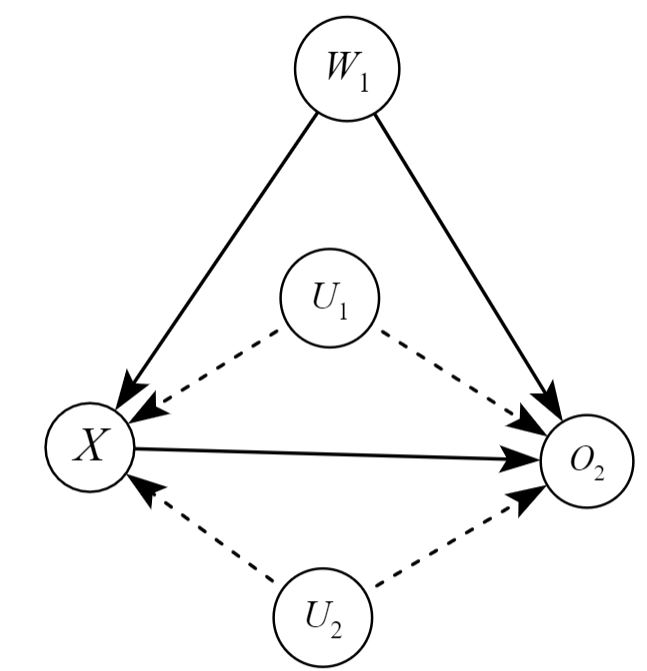}
  \subcaption{}
\end{subfigure}%
\begin{subfigure}{.6\textwidth}
  \centering
  \includegraphics[width=1\linewidth]{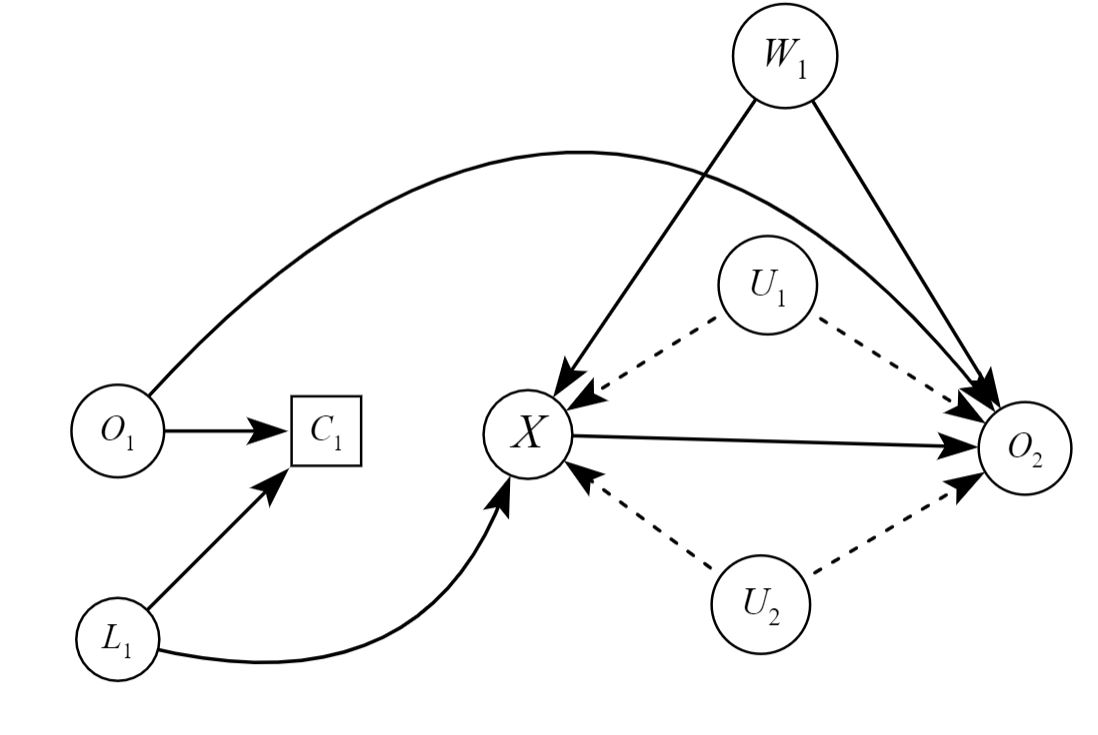}
   \subcaption{}
\end{subfigure}
\caption{DAGs for a pretest-posttest Design}
\label{pre-post2}
\end{figure}
\\ 
\subsection{\textbf{Designs with a control group but no pretest}} \par
3. \textbf{Posttest-Only Design With Nonequivalent Groups:} \par
     \hspace{3cm} NR \hspace{2cm}   X      \quad \quad $O_1$ \par
     \hspace{3cm} NR \hspace{2cm}    \quad      \quad \quad   $O_1$ \par 
This is similar to a one-group posttest design with the difference of adding a control group. The control group is supposed to provide a counterfactual outcome for the treated group had they not received the treatment. The DAG for this is demonstrated in figure \ref{posttest_nonequivalent} and is similar to the DAG in figure \ref{posttest} with the inclusion of variables $Z$. $Z$ is a variable indicating the treatment assignment. It determines the conditions on how to assign the participants to treatment and control groups, such as choosing some classrooms for the treatment and some others for the control group. One major problem with the nonequivalent design is the selection bias since the participants are not assigned randomly. Therefore, the participants in the control and treatment groups do not share identical characteristics. For example, assume participants in treatment groups are from more educated family backgrounds, or they are academically stronger candidates. Therefore they perform better in the posttest even under the null treatment effect. 
\begin{figure}[h]
    \centering
    \includegraphics[width=0.6\textwidth]{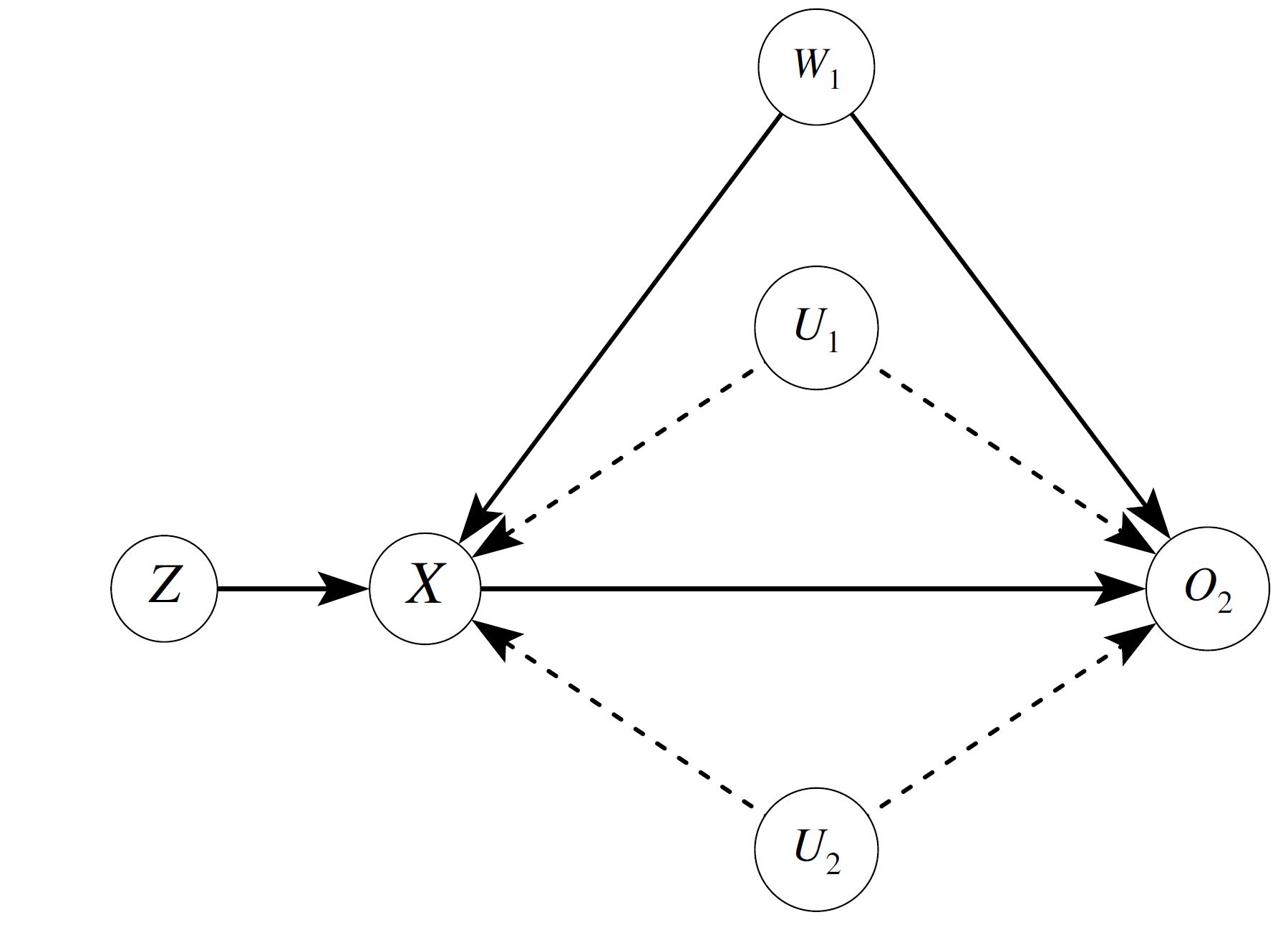}
    \caption{The DAG for Posttest-Only Design With Nonequivalent Groups}
    \label{posttest_nonequivalent}
\end{figure}
\subsection{Designs with both control groups and pretests} \par
4. \textbf{The Pretest-Posttest Design with Nonequivalent Groups:} \par
     \hspace{3cm} NR \hspace{2cm}$O_1$ \quad \quad X      \quad \quad $O_2$ \par
     \hspace{3cm} NR \hspace{2cm}$O_1$   \quad\quad \quad      \quad \quad   $O_2$ \par 
This is a common design in the educational systems. Students are divided into treatment and control groups based on some factors, such as the classroom. The DAG after the intervention for this design is similar to figure \ref{posttest_nonequivalent}. 
The control and pretest provide a counterfactual for the treated group and help to remove some of the threats to the internal validities. However some flaws with this design are as follows: 1- A single intervention may not be sufficient to influence the outcome (posttest). 2- Since the participants are not randomly assigned, the treatment and control groups may have very different characteristics (confounder). 3- A single observation before and after the intervention is not enough to draw any conclusion about the participants' mastery of the subject and is prone to regression to the mean threats. 4- Since there is a single intervention then there is no opportunity to provide feedback to students.
\\
\subsection{Design for time series data} 
5. \textbf{Interrupted Time series Design} \par 
\hspace{3cm} $O_1$ \hspace{1cm} $O_2$ \hspace{1cm} ... \hspace{1cm} $O_n$ \hspace{1cm}  X  \hspace{1cm} $O_{n+1}$ \hspace{1cm} . . . \hspace{1cm}  $O_{2n}$\par
Interrupted time series (ITS) design consists of repeated measurements over time with an intervention, or a shock is introduced at a certain time. It is considered the most powerful design in the hierarchy of quasi-experimental designs. This is because it can rule out many threats to internal validity such as regression to the mean since many measurements are taken before and after the intervention; therefore, the probability that the result is only due to chance is very slim. However, history and instrumentation are still considered as treats to internal validities since due to administrative change, the measure for student learning can change through time. Attrition is a big threat for this design since the loss-to-follow-up can happen after each measurement. Therefore the number of participants from the beginning to the end of the study can be very different.
Figure \ref{ITS1} shows a candidate DAG for this design for four measurements (2-time points before and 2-time points after the intervention). The $O_1$, $O_2$, . . ., $O_4$ denote the measurements, $L_1$, $L_2$, . . ., $L_4$ are the time-varying covariates such as family, financial or prerequisite problems that influence the outcome and treatment. It is important to note that figure \ref{ITS1} is a simplified DAG for most real-world phenomena. 
\begin{figure}[h]
    \centering
    \includegraphics[width=1\textwidth]{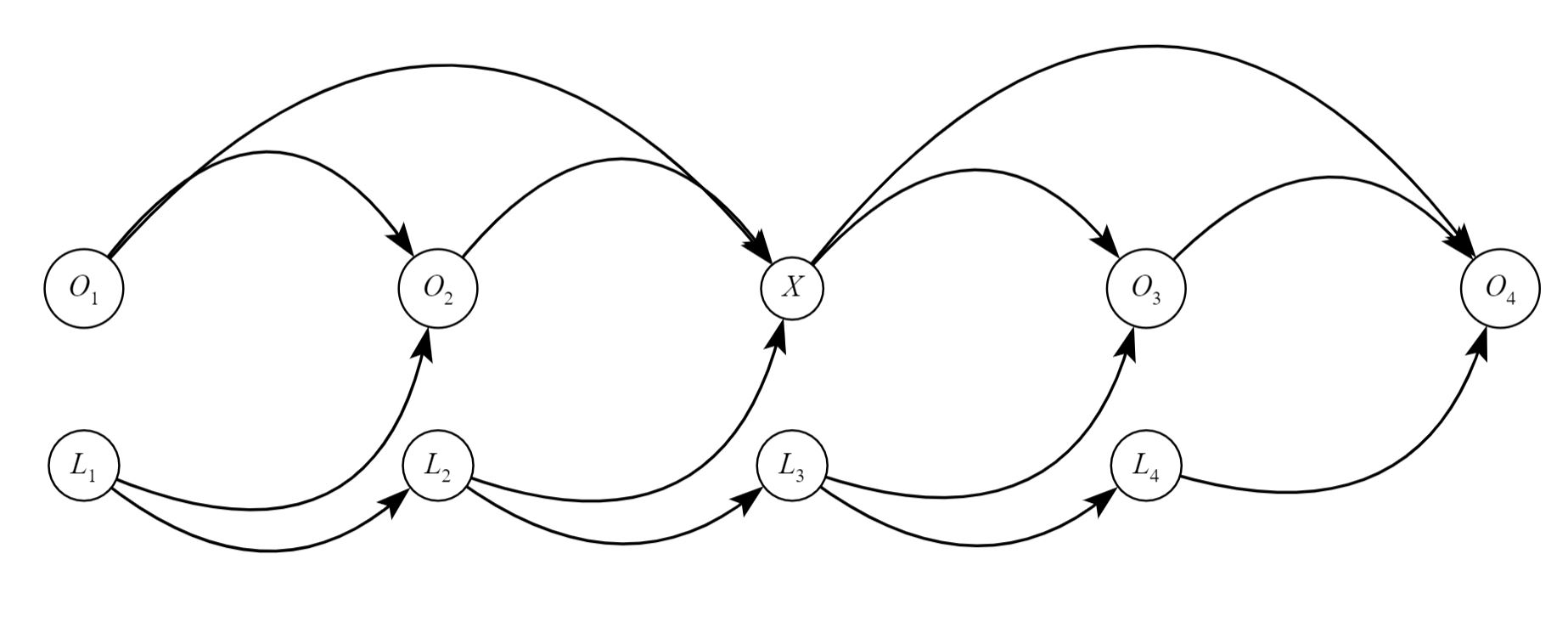}
    \caption{A DAG for the Interrupted Time Series Design}
    \label{ITS1}
\end{figure}
\\
Interrupted time series (ITS) is a type of design used when a randomized experiment is either too expensive or impractical. ITS measures the change in intercept and slope of the time series before and after the intervention. It comprises two distinct periods, namely the pre-intervention and the post-intervention, and may include the control group. There are two major methods to model the ITS design. The segmented regression by far is the most widely used one, in which a linear regression model is fit for before and after intervention with a dummy variable representing the pre and post-intervention phase. This model assumes a linear trend before and after intervention time \cite{turner2020evaluation}, \cite{wagner2002segmented},and \cite{taljaard2014use}. The segmented regression technique assumes the residual term to be normally distributed, but if residuals are autocorrelated, then autoregressive integrated moving average (ARIMA) can be used to model the ITS \cite{turner2020evaluation}, \cite{schaffer2021interrupted}, and \cite{gottman1978analysis}. This method works by first building the ARIMA model of the time series using only the pre-intervention series. Then apply the model for the complete time series data and add the intervention as a dummy variable, where 1 is coded as the post-intervention phase, and 0 is for the baseline period. ARIMA only handles the univariate time series data.\\
ITS can be used in education to evaluate the influence of a single intervention introduced in a semester as a form of a study skill or measure the influence of adding an online textbook.
\subsection{Regression Discontinuity (RD)}\par
\hspace{2cm} \textcolor{blue}{C} \hspace{1cm}$O_1$ \quad \quad X      \quad \quad $O_2$ \par
\hspace{2cm} \textcolor{blue}{C} \hspace{1cm}$O_1$   \quad\quad      \quad \quad \quad $O_2$     \par 
RD is another type of quasi-experimental design used when there is some kind of criterion (threshold) that must be met before participating in the intervention. This threshold identifies the eligibility for participation in the program.\\ It is a type of pretest-posttest comparison group, where participants are assigned to treatment and comparison groups based on the threshold or cutoff on a pretest. This can be represented mathematically as follows: $A$ is called the running variable and $C$ is the threshold. If $A$ is above the threshold then you receive the intervention. \\
\[
X =
        \begin{cases}
          1   & A\geq C \\
          0   & A < C    
        \end{cases}
        \]
Figure \ref{RD1} represents a candidate DAG for RD design. According to this figure, $A$, $L$, and $U$ are confounders ($U$ is an unobserved confounder). and  $X\leftarrow A\leftarrow L\rightarrow O$, $X\leftarrow A\leftarrow U\rightarrow O$, and $X\leftarrow A\rightarrow O$ are the backdoor paths. Therefore, one way to obtain the unbiased estimate of $X$ on $O$ is to control for $A$. In RD, we consider the effect of the intervention on the outcome when the running variable ($A$) is near the threshold (condition on $A$). Hence, by doing this, we will close all the backdoor paths. \\
In education, $A$ can be a pretest, $X$ is the intervention or treatment such as talking to a TA or a tutor, $O$ is the posttest, $L$ is the observed confounder such as the prerequisite knowledge, $U$ is the unobserved confounder such as family background, and $C$ is the threshold on the pretest for participation in the program. In RD, we study the influence of the intervention on the posttest score for those participants whose pretest score is near the threshold. \\
To analyze the RD for the above problem in education, let $A$ denote the pretest score,$O$ denote the posttest score, and $Z$ be a binary dummy variable indicating the treatment or control group. Then select the participants whose pretest score is just below and above the threshold, and regress the posttest score on the pretest and other variables.
\begin{align}
        O = \beta_0 + \beta_1A + \beta_2Z + \beta_3 AZ + \epsilon
        \label{RD2}
    \end{align}
The $\beta_3$ coefficient is the causal effect of $A$ on $O$.
\begin{figure}[h]
    \centering
    \includegraphics[width=0.4\textwidth]{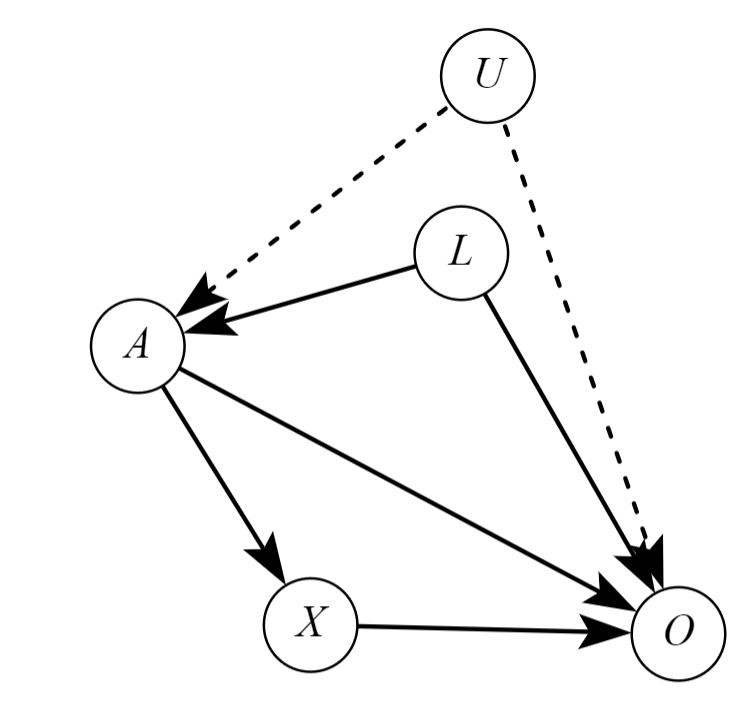}
    \caption{A DAG for Regression Discontinuity Design}
    \label{RD1}
\end{figure}
\\
Other popular quasi-experimental designs are Difference-in-differences (DID) and instrumental variable (IV) that can provide unbiased causal estimates when a randomized experiment is not possible \cite{branas2011difference},\cite{donald2007inference}, \cite{dimick2014methods},\cite{martens2006instrumental}, \cite{angrist1996identification}, and \cite{greenland2000introduction}.\\
Although quasi-experimental designs such as ITS, RD, DID, and IV provide unbiased causal estimates for many real-world phenomena, they may not be suitable options for education since 1- They mainly model a single intervention yet, learning is more effective using a sequence of interventions. 2- They do not model time-varying treatment and confounder feedback; therefore, they do not incorporate any mechanisms to reduce the negative effect of confounders through timely intervention. \\
Our objective is to have a tractable approach for the actual educational system by accounting for all the potential confounders and design suitable interventions to reduce the negative influence of confounders through time. Thus, in the next section, we discuss observational studies in education and introduce techniques to overcome  limitations of quasi-experimental designs with a single intervention.
\section{Non-randomized Designs with Multiple Interventions in Education}
Before discussing non-randomized designs with time varying interventions, We first define parameters and notation used in this section:
\begin{itemize}
    \item $X_1$, $X_2$ , . . . $X_T$ are time varying treatments, such as:
    \begin{itemize}
        \item Reading chapter of book.
        \item Solving some problems.
        \item Talking to the TA/instructor/counselor.
    \end{itemize}
    \item $L_1$, $L_2$ , . . . $L_T$ are observed and $U_1$, $U_2$ , . . . $U_T$ are unobserved time varying confounders, such as:
     \begin{itemize}
        \item Prior knowledge of subject.
        \item Psychological/stress problems.
    \end{itemize}
    \item $O$ is the outcome. such as exam score.
    \item $N$ is the total number of students.
    \end{itemize}
The \textbf{ Objective} is to find the causal effect of joint interventions $X_1$,$X_2$, . . . , $X_T$ on $O$. The Joint effect of interventions is defined as a collection of direct effects unmediated by other interventions \cite{daniel2013methods}. For example, joint effect of interventions $X_1$ and $X_2$ on $O$ in figure \ref{joint_effect1} consists of $X_1 \rightarrow O$ and $X_2\rightarrow O$. In this section, we consider the following three cases, which characterize scenarios happening in real-world educational systems.
    \begin{figure}[tbh]
        \centering
        \includegraphics[scale=0.55]{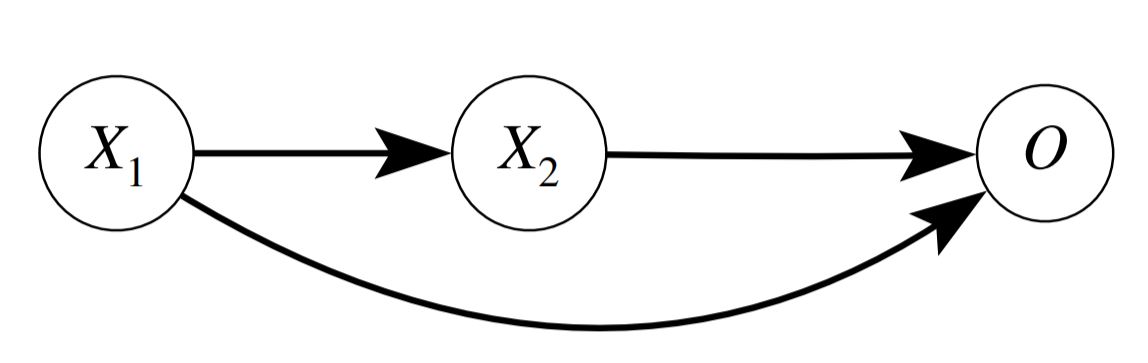}
        \caption{Joint Effect of $X_1$ and $X_2$ on $O$}
        \label{joint_effect1}
    \end{figure}
\begin{enumerate}
    \item Time-Varying Treatment and Confounder with no unmeasured confounders.
    \item Time-Varying Treatment and Confounder with unmeasured confounders.
    \item Time-Varying Treatment-Confounder feedback.
\end{enumerate}
\subsection{Time Varying Treatments and Confounders With No Unmeasured Confounders}
Figure \ref{Time_varying_Treatment_NU} presents an example of this scenario, where $X_1$, $X_2$ and $X_3$ are referred to as time-varying treatments/interventions. They can represent solving problems, talking to the TA, and talking to the counselor. $L_1$, $L_2$, and $L_3$ are observed confounders such as student background knowledge through time. $O$ is the posttest exam. We use the outcome regression model to find the average causal effect of joint intervention $X_1$,$X_2$, and $X_3$ on $O$. The outcome regression measures the correlation coefficients but not necessarily the causal coefficients. Therefore we need to investigate if the correlation and causation are the same in equation \ref{outcome_regression1}. Controlling for $L_1$, $L_2$ and $L_3$ will close all the backdoors such as  $X_1 \leftarrow L_1\rightarrow O$, $X_2 \leftarrow L_2\rightarrow O$, and $X_3 \leftarrow L_3 \rightarrow O$. Furthermore controlling for $X_1$, $X_2$ and $X_3$ will block the mediated and all other backdoor paths such as $X_2 \leftarrow X_1 \rightarrow O$. Hence the outcome regression in equation \ref{outcome_regression1} results in unbiased estimates of causal coefficients. In equation \ref{outcome_regression1}, the multiplication terms are interaction terms, for example $X_1X_2$ suggests that the causal effect of $X_1$ on $O$ is different for different values of $X_2$. 
\begin{figure}[tbh]
    \centering
    \includegraphics[scale=0.80]{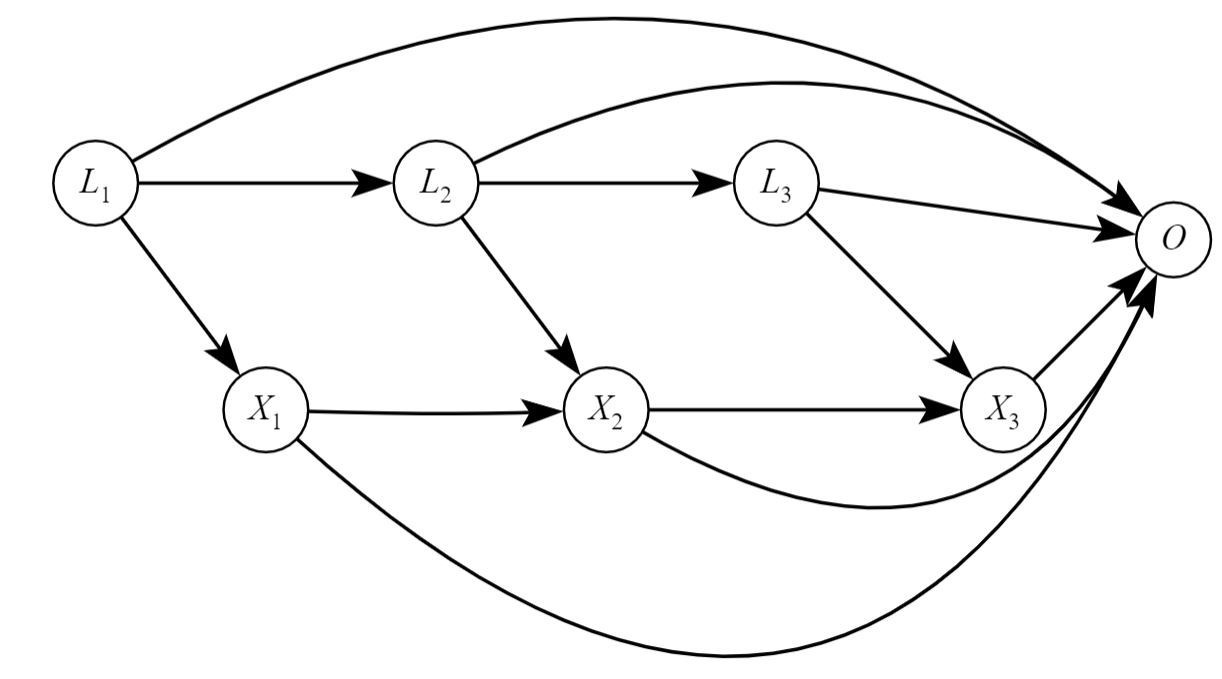}
    \caption{Time Varying Treatment and Confounder for three time points}
    \label{Time_varying_Treatment_NU}
\end{figure}
\begin{align}
    E[O|X_1,X_2,X_3,L_1,L_2,L_3]  = \alpha_0 + \alpha_1 X_1 + \alpha_2 X_2 + \alpha_3 X_3 + \alpha_4 X_1X_2 + \nonumber \\ \alpha_5 X_1X_3 + \alpha_6 X_2X_3 + \alpha_7 X_1X_2X_3 + \alpha_8 L_1 + \alpha_9 L_2 + \alpha_{10} L_3 
    \label{outcome_regression1}
\end{align}
The following is a simulation setup for this case:
\begin{align*}
    N=& 500 \\
    P(L_1 = 1) =& 0.5 \\
    P(X_1 = 1) =& 0.2 + 0.4L_1 \\
    P(L_2 = 1) =& 0.2+0.6L_1 \\
    P(X_2 = 1) =& 0.3 + 0.5X_1+0.2L_2 \\
    P(L_3 =1) =& 0.3+0.5L_2 \\
    P(X_3=1) =& 0.3 + 0.4X_2+ 0.1L_3 \\
    \mu =& 0.2 + 2X_1 + 6X_3 + 5X_2+ X_1X_2+ X_1X_3+ X_2X_3 + 4L_1 + 4L_2+3L_3 \\
    \sigma =& 0.2 \quad \quad O \sim N(\mu,\sigma^2)\\
\end{align*}
$N$ represents number of cases, confounders and treatments are all binary variables and the posttest is normally distributed. Table \ref{LOC1} shows the simulation result for this scenario. It indicates all the coefficients except $X_1X_2X_3$ are significant for P-value of $0.05$. Therefore the coefficient of $X_1X_2X_3$ may be assumed to be zero and the regression model estimates all coefficients correctly. Table \ref{LOC1} only includes the coefficients for treatments and their interaction terms and not any information about the confounders. Table \ref{Bootstrap_regression1} shows the $95\%$ confidence interval (CI) for the coefficients using bootstrapping. According to this table the true values for all the coefficients lie within the $95\%$ confidence interval. 
\begin{table}[tbh]
\centering
    \caption{Simulation Result for Time-Varying Treatments\\, Confounders with No Unmeasured Confounders}
    \begin{tabular}{|c|c|c|c|}
    \hline
         Coefficient & True & Estimate & P-Value  \\
         \hline
            Intercept & $0.2$ & $0.17$ & $0.0008$ \\
            \hline
            $X_1$ &  $2$   &  $1.90$ & $2 x 10^{-16}$\\
            \hline
            $X_2$ &  $5$   &  $4.98$ & $2 x 10^{-16}$\\
            \hline
            $X_3$ &   $6$  &  $6.02$ & $2 x 10^{-16}$  \\
            \hline
            $X_1X_2$ &  $1$  & $0.91$ & $2 x 10^{-7}$ \\
            \hline
            $X_1X_3$ &  $1$  &  $1.34$ & $9.6 x 10^{-7}$\\
            \hline
            $X_2X_3$ &  $1$  &  $0.90$ & $1.9 x 10^{-7}$\\
            \hline
            $X_1X_2X_3$ &  $0$  &  $-0.091$ & $0.81$\\
            \hline
    \end{tabular}
    \label{LOC1}
\end{table}

\begin{table}[tbh]
\centering
    \caption{Simulation Result for Time-Varying Treatments\\, Confounders with No Unmeasured Confounders}
    \begin{tabular}{|c|c|c|c|}
    \hline
         Coefficient & True & $95\%$ CI \\
         \hline
            Intercept & $0.2$ & $(0.0702, 0.2826)$ \\
            \hline
            $X_1$ &  $2$   & $(1.710, 2.302)$\\
            \hline
            $X_2$ &  $5$   &  $(4.875, 5.336)$\\
            \hline
            $X_3$ &   $6$  &   $(5.826, 6.180)$  \\
            \hline
            $X_1X_2$ &  $1$  & $ (0.4688, 1.2230)$ \\
            \hline
            $X_1X_3$ &  $1$  &  $(0.755, 1.918)$\\
            \hline
            $X_2X_3$ &  $1$  &  $(0.5542, 1.1484)$\\
            \hline
            $X_1X_2X_3$ &  $0$  &  $(-0.8461, 0.4669)$\\
            \hline
    \end{tabular}
    \label{Bootstrap_regression1}
\end{table}
The results in this section suggest that if all the confounders are measured properly, then the unbiased causal effect of joint interventions on the outcome can be estimated accurately. Next, we investigate what would happen if there are unmeasured confounders.
\subsection{Time Varying Treatments and Confounders With Unmeasured Confounders}
Figure \ref{Time_varying_Treatment_WU} presents an example of this scenario, where $X_1$, $X_2$ and $X_3$ are referred to as time varying treatments/interventions. Similar to the figure \ref{Time_varying_Treatment_NU} interventions can represent solving problems, talking to the TA, and talking to the counselor. $L_1$, $L_2$ and $L_3$ are observed confounders such as student background knowledge through time. $U_1$, $U_2$ and $U_3$ are unobserved confounders such as family and psychological problems. $O$ is the posttest exam. To find the average causal effect of joint intervention $X_1$,$X_2$, and $X_3$ on $O$, we use the same regression model as in previous case with all observed confounders. We need to investigate if the correlation and causation are the same in equation \ref{outcome_regression2}. Controlling for $L_1$, $L_2$ and $L_3$ will close backdoors such as  $X_1 \leftarrow L_1\rightarrow O$, $X_2 \leftarrow L_2\rightarrow O$, and $X_3 \leftarrow L_3 \rightarrow O$. Furthermore controlling for $X_1$, $X_2$ and $X_3$ will block the mediated and all other backdoor paths such as $X_1 \leftarrow X_2 \rightarrow O$. However any backdoors through unobserved confounders such as $X_1 \leftarrow U_1 \rightarrow O$ and $X_2 \leftarrow U_2 \rightarrow O$ remain open. Hence the outcome regression in equation \ref{outcome_regression2} results in biased estimate of causal coefficients.
Note equations \ref{outcome_regression1} and \ref{outcome_regression2} are the same since unobserved variables cannot be included in the regression model.
\begin{figure}[tbh]
    \centering
    \includegraphics[scale=0.80]{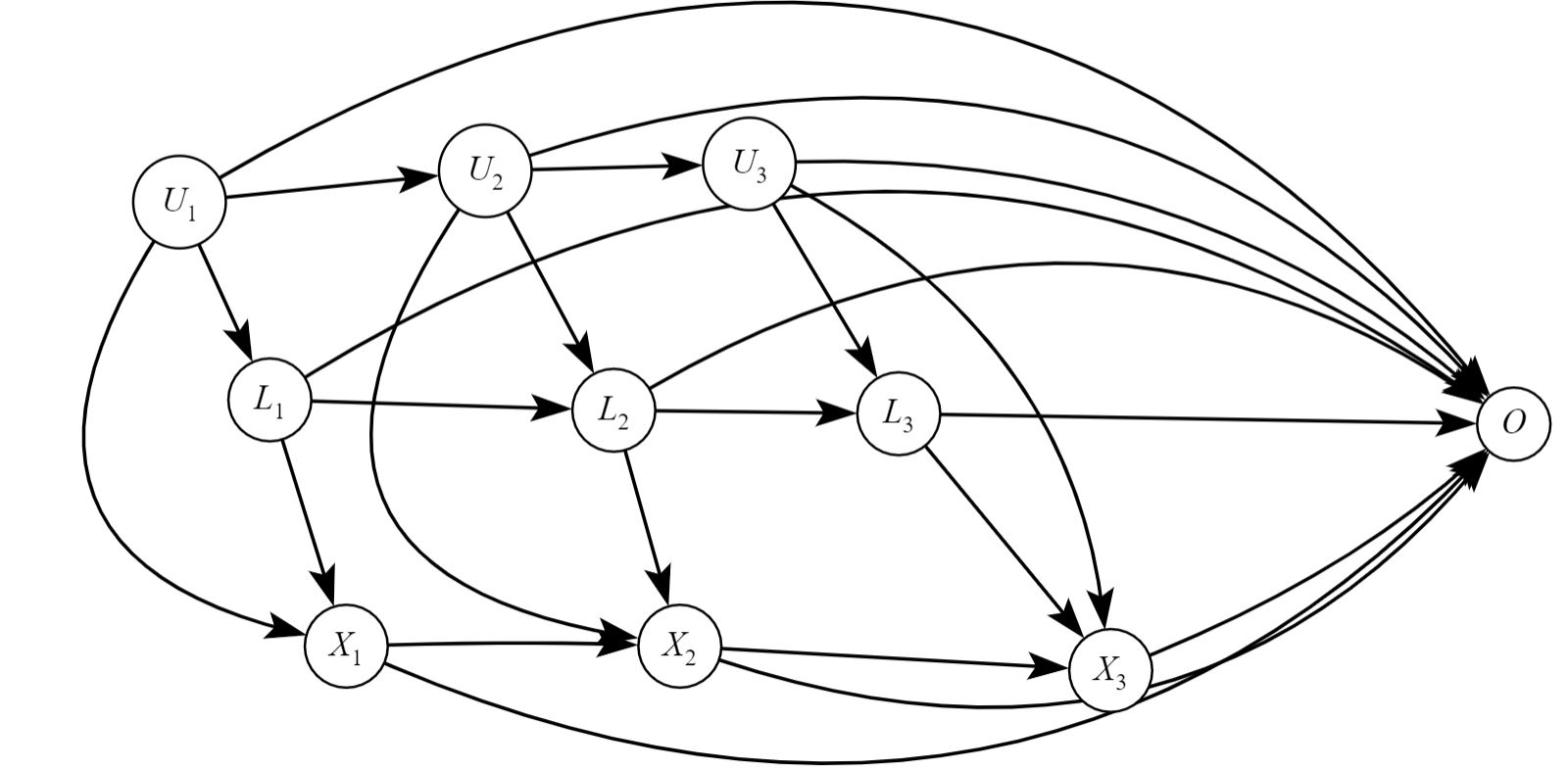}
    \caption{Time Varying Treatment and Confounder for three time points \\ with Unobserved Confounders}
    \label{Time_varying_Treatment_WU}
\end{figure}
\begin{align}
    E[O|X_1,X_2,X_3,L_1,L_2,L_3]  = \alpha_0 + \alpha_1 X_1 + \alpha_2 X_2 + \alpha_3 X_3 + \alpha_4 X_1X_2 + \nonumber \\ \alpha_5 X_1X_3 + \alpha_6 X_2X_3 + \alpha_7 X_1X_2X_3 + \alpha_8 L_1 + \alpha_9 L_2 + \alpha_{10} L_3 
    \label{outcome_regression2}
\end{align}
The following is a simulation setup for this case:
\begin{align*}
    &N= 500 \\
    &P(U_1 = 1) =  0.6           &    &P(U_2 = 1) = 0.6 + 0.3U_1 \\
    &P(U_3 = 1) = 0.4 + 0.3U_2  &    &P(L_1 = 1) = 0.5 + 0.4U_1 \\
    &P(X_1 = 1) = 0.2 + 0.4L_1 + 0.3U_1  &  &P(L_2 = 1) = 0.2+0.6L_1 + 0.1U_2 \\
    &P(X_2 = 1) = 0.3 + 0.2X_1+0.2L_2 + 0.2U_2    & &P(L_3 =1) = 0.3+0.5L_2 + 0.1U_3 \\
    &P(X_3=1) = 0.1 + 0.4X_2+ 0.2L_3 + 0.2U_3  \\
    &\mu=  4L_1 + 4L_2+3L_3+ 5X_2 + 2X_1 + 6X_3 + X_1X_2 \\ 
    &X_1X_3+ X_2X_3+ 2U_1 + 3U_2 + U_3 + 0.2 \\
    &\sigma = 0.2 \hspace{1cm}  O \sim N(\mu,\sigma^2)\\
\end{align*}
Table \ref{UUC1} shows the simulation result for this case. According to this table the coefficients of $X_1X_2$, $X_1X_3$, $X_2X_3$, and $X_1X_2X_3$ are not statistically significant for P-value of $0.05$. This is not correct since we know the coefficients of $X_1X_2$, $X_1X_3$, $X_2X_3$ should not be zero. Table \ref{Bootstrap_regression2} shows the $95\%$ confidence interval (CI) for the coefficients using bootstrapping. According to this table, the $95\%$ confidence interval for the intercept, $X_2$ and $X_3$ do not contain the true values. Moreover, the confidence interval, in this case, is wider than the confidence interval when all the confounders are observed. 
\begin{table}[tbh]
\centering
    \caption{Simulation Result for Time-Varying Treatments\\, Confounders with Unmeasured Confounders}
    \begin{tabular}{|c|c|c|c|}
    \hline
         Coefficient & True & Estimate & P-Value  \\
         \hline
            Intercept & $0.2$ & $1.92$ & $10^{-11}$ \\
            \hline
            $X_1$ &  $2$   &  $2.97$ & $2 x 10^{-11}$\\
            \hline
            $X_2$ &  $5$   &  $6.15$ & $2 x 10^{-16}$\\
            \hline
            $X_3$ &   $6$  &  $7.01$ & $2 x 10^{-16}$  \\
            \hline
            $X_1X_2$ &  $1$  & $1.24$ & $0.056$ \\
            \hline
            $X_1X_3$ &  $1$  &  $0.55$ & $0.42$\\
            \hline
            $X_2X_3$ &  $1$  &  $0.16$ & $0.80$\\
            \hline
            $X_1X_2X_3$ &  $0$  &  $0.37$ & $0.67$\\
            \hline
    \end{tabular}
    \label{UUC1}
\end{table}

\begin{table}[tbh]
\centering
    \caption{Simulation Result for Time-Varying Treatments\\, Confounders with Unmeasured Confounders}
    \begin{tabular}{|c|c|c|c|}
    \hline
         Coefficient & True & $95\%$ CI \\
         \hline
            Intercept & $0.2$ & $( 1.844,  2.967 )$ \\
            \hline
            $X_1$ &  $2$   & $( 1.657,  3.638 )$\\
            \hline
            $X_2$ &  $5$   &  $( 5.19,  7.062 )$\\
            \hline
            $X_3$ &   $6$  &   $( 5.486,  8.419 )$  \\
            \hline
            $X_1X_2$ &  $1$  & $ (-0.828,  3.084 )$ \\
            \hline
            $X_1X_3$ &  $1$  &  $ (-0.212,  3.917 )$\\
            \hline
            $X_2X_3$ &  $1$  &  $(-1.4414,  2.6802 )$\\
            \hline
            $X_1X_2X_3$ &  $0$  &  $(-3.3632,  1.8823 )$\\
            \hline
    \end{tabular}
    \label{Bootstrap_regression2}
\end{table}
The accuracy of the estimated coefficient depends on the influence of unobserved confounders. As an example, consider the following two case studies where the simulation setup is the same as before except the mean of the outcome $\mu$ varies as a function of unobserved confounders. \\
Case I :\textbf{Unobserved confounders have small influence on the outcome}:
\begin{align*}
    \mu_1 &=  4L_1 + 4L_2+3L_3+ 5X_2 + 2X_1 + 6X_3 + X_1X_2 +
    X_1X_3+ X_2X_3+ 0.1U_1 + 0.2U_2 + 0.3U_3 + 0.2 \\
    \sigma_1 &= 0.2 \\
    O_1 &\sim N(\mu_1,\sigma_1^2)
\end{align*}
Case II :\textbf{Unobserved confounders have huge influence on the outcome}:
\begin{align*}
    \mu_2 &=  4L_1 + 4L_2+3L_3+ 5X_2 + 2X_1 + 6X_3 + X_1X_2 +
    X_1X_3+ X_2X_3+ 8U_1 + 9U_2 + 10U_3 + 0.2 \\
    \sigma_2 &= 0.2 \\
    O_2 &\sim N(\mu_2,\sigma_2^2)
\end{align*}
 Table \ref{UUC2} summarizes the result for these two case studies and clearly demonstrates the negative effect of unobserved confounders on the estimated coefficients. Under case II, the estimated coefficients are far from the true coefficients, and this is due to the large influence of open backdoor paths.
\begin{table}[tbh]
\centering
    \caption{Simulation Result for Time-Varying Treatments, Confounders with Unmeasured Confounders (Comparison of Case I and II)}
    \begin{tabular}{|c|c|c|c|}
    \hline
         Coefficient & True & Estimate (Case I) & Estimate (Case II)  \\
         \hline
            Intercept & $0.2$ & $0.41$ & $8.97$ \\
            \hline
            $X_1$ &  $2$   &  $2.08$ & $7.17$\\
            \hline
            $X_2$ &  $5$   &  $5.12$ & $7.91$\\
            \hline
            $X_3$ &   $6$  &  $5.84$ & $10.05$  \\
            \hline
            $X_1X_2$ &  $1$  & $1.22$ & $0.64$ \\
            \hline
            $X_1X_3$ &  $1$  &  $1.34$ & $1.98$\\
            \hline
            $X_2X_3$ &  $1$  &  $1.84$ & $0.04$\\
            \hline
            $X_1X_2X_3$ &  $0$  &  $0.44$ & $-1.58$\\
            \hline
    \end{tabular}
    \label{UUC2}
\end{table}
\\
The results in this section suggest that open backdoor paths due to unobserved confounders can cause bias in the estimation of the causal coefficients. The amount of bias depends on how large is the influence of the unobserved confounders. This is another important conclusion in designing more intelligent and individualized educational systems that we need to measure and control for as many confounders as we can. It is true that no matter how many confounding variables we measure, there are still some unobserved confounders; however, if we can close for the confounding variables that might plausibly have a large influence on the outcome, then we reduce the bias in the causal coefficients significantly (refer to the case I and case II). In the next section, we discuss time-varying treatments, confounders feedback.
\subsection{Time Varying Treatments and Confounders Feedback}
Figure \ref{Feedback_iptw} presents an example of this scenario, where $X_1$, $X_2$ and $X_3$ are referred to as time-varying treatments/interventions. Similar to the previous two sections. interventions can represent solving problems, talking to the TA, and talking to the counselor. $L_1$, $L_2$, and $L_3$ are observed confounders such as student background knowledge through time. $O$ is the posttest exam. We assume we control for a sufficient set of confounders, and there are no significant unobserved confounders. The difference between this DAG and the DAG in figure \ref{Time_varying_Treatment_NU} is that there are arrows from treatments back to confounders. The motivation behind this design is as follows: assume some students are suffering from a psychological problem, we refer them to the counselor and are hoping through the counselor intervention we reduce the negative effect of the confounder. Therefore a more intelligent design requires feedback from treatments to interventions. To find the average causal effect of joint intervention $X_1$,$X_2$, and $X_3$ on $O$, we start again by the same regression model as in previous cases with all observed confounders. We need to investigate if the correlation and causation are the same in equation \ref{iptw_1} as in other cases. Controlling for $L_1$, $L_2$ and $L_3$ will close backdoors such as  $X_1 \leftarrow L_1\rightarrow O$, $X_2 \leftarrow L_2\rightarrow O$, and $X_3 \leftarrow L_3 \rightarrow O$. Furthermore controlling for $X_1$, $X_2$ and $X_3$ will block the mediated and all other backdoor paths such as $X_2 \leftarrow X_1 \rightarrow O$. However controlling for $L_2$ will close the $X_1 \rightarrow L_2 \rightarrow O$ path, which is a causal path (neither mediated nor a backdoor paths), the same way controlling for $L_3$, will block the causal path $X_2 \rightarrow L_3 \rightarrow O$. Therefore we conclude that the coefficients of $X_2$ and $X_3$ are biased and outcome regression produces an unbiased coefficient for $X_1$. \\ 
\begin{figure}[tbh]
    \centering
    \includegraphics[scale=0.80]{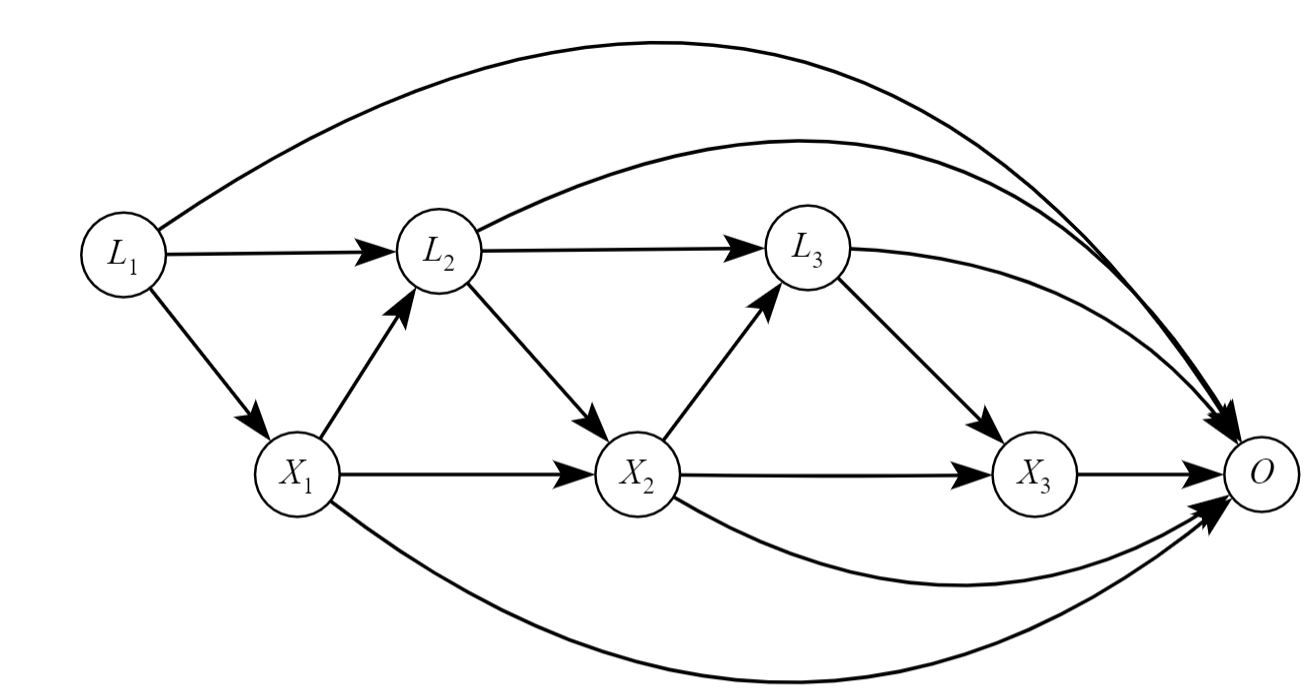}
    \caption{Time Varying Treatments and Confounders Feedback}
    \label{Feedback_iptw}
\end{figure}
\begin{align}
    E[O|X_1,X_2,X_3,L_1,L_2,L_3]  = \alpha_0 + \alpha_1 X_1 + \alpha_2 X_2 + \alpha_3 X_3 + \alpha_4 X_1X_2 + \nonumber \\ \alpha_5 X_1X_3 + \alpha_6 X_2X_3 + \alpha_7 X_1X_2X_3 + \alpha_8 L_1 + \alpha_9 L_2 + \alpha_{10} L_3 
    \label{iptw_1}
\end{align}
To find the unbiased causal estimates of joint interventions on the outcome in the case of time-varying treatments, confounders feedback, we use g-formula and the inverse probability of treatment weighting (IPTW). Before discussing how to do inference using g-formula and IPTW, we will briefly discuss each one in the next section.
\subsubsection{G-formula and Inverse Probability of Treatment Weighting}
G-formula is an extended version of the backdoor criterion or standardization discussed before in the time-varying treatments and confounders setting. Mathematically it is represented as follows:
\begin{align}
E[O_{\bar{x}}] = \sum_{\bar{l}} E[O|\bar{X} = \bar{x}, \bar{L} = \bar{l}] \prod_{t=0}^{t=T} P(L_t = l_t | \bar{X}_{t-1} = \bar{x}_{t-1}, \bar{L}_{t-1} = \bar{l}_{t-1})
\end{align}
where the $\bar{X}$ and $\bar{L}$ represent the treatments and confounders history, $T$ is the number of time steps or the number of interventions. This model requires estimating the conditional expectation and conditional probability from data. Hence it is sensitive to model misspecification and requires correct models for conditional expectation and conditional probabilities.\\ 
IPTW is a method to weigh each observation by the inverse of the probability of treatment received. IPTW creates what is called the pseudo-population, which is a confounder-free population. Therefore IPTW is trying to emulate a randomized experiment. 
The IPTW weights for the subject i is defined as follows \cite{daniel2013methods}:
\begin{align}
    W_i &= \dfrac{1}{\prod_{t=0}^{t=T} P_{X_{t}|\bar X_{t-1},\bar L_t}(X_{t,i}, \bar X_{t-1,i}, \bar L_t,i)} \\
    SW_i &= \dfrac{\prod_{t=0}^{t=T} P_{X_t|\bar X_{t-1}} (X_{t,i}|X_{t-1,i})}{\prod_{t=0}^{t=T} P_{X_{t,i}|\bar X_{t-1,i},\bar{L_t}}(X_{t,i}, \bar X_{t-1,i}, \bar{L_t,i})} 
\end{align}
The $SW_i$ is referred to as the stabilized weight. It results in a more stable estimator if the denominator of $W_i$ is close to zero. 
However, a better way to express the IPTW weights is through a causal DAG. 
The denominator of IPTW weights (It is the same for $W_i$ and $SW_i$) is the probability distribution of the treatments in the pre-intervention DAG since the treatment at the current time point is conditioned on treatment history and the confounder history.  The numerator of the $SW$ weight is the probability distribution on the post-intervention DAG since it is independent of confounders; hence the connections from confounders to treatments are removed. Therefore the stabilized weight can be represented as follows:
\begin{align}
    IPTW_{SW} = \dfrac{\textrm{Distribution of Treatment on Post-intervention DAG}}{\textrm{Distribution of Treatment on pre-intervention DAG}}  
    \label{sw_eq1}
\end{align}
IPTW is used to estimate the parameters of marginal structural model (MSM) \cite{robins2000marginal}, \cite{vanderweele2009marginal}, and \cite{cole2008constructing}. MSM, as the name suggests, is a marginal model, namely not conditioned on any confounding variables. It is a model for calculating the mean potential outcome for the entire population. To see how g-formula and IPTW can solve the time-varying treatments and confounders feedback problems, consider the following simulation setup corresponding to the DAG in figure \ref{Feedback_iptw} with the difference that we assume the first treatment $X_1$ is assigned randomly; hence there is no confounder connected to the $X_1$:
\begin{align*}
    &N= 500 \\
    &P(X_1=1) = 0.6 \\
    &P(L_2 = 1) = 0.1+0.6X_1 \\
    &P(X_2 = 1) = 0.3 + 0.7L_1 - 0.25X_1 \\
    &P(L_3 = 1) = 0.2 + 0.3L_2 + 0.3X_2 \\
    &P(X_3 = 1) = 0.1 + 0.3L_3  + 0.1X_2   \\
    &\mu=  5+ 4X_1 + 5X_2 + 6X_3 + 10X_1X_2 + 11X_1X_3 + 12X_2X_3 - 8L_1 - 9L_2\\ 
    &\sigma = 0.2 \hspace{1cm}  O \sim N(\mu,\sigma^2)\\
\end{align*}
As it was discussed, the coefficients of $X_2$ and $X_3$ are not causal; therefore, to find the true causal coefficients we proceed as follows:
\begin{align}
    &E[Y_{X_1X_2X_3}] = \sum_{L_2} \sum_{L_3} E[Y_{X_1X_2X_3}|L_2,L_3] = \sum_{L_2} \sum_{L_3} 5+ 4X_1 + 5X_2 + 6X_3 + 10X_1X_2 + 11X_1X_3+  \\& 12X_2X_3 -8L_1 - 9L_2 \nonumber \\
     &E[Y_{X_1X_2X_3}] = 2.13 -2.42X_1 + 2.3 X_2 + 6X_3 + 10X_1X_2 + 11X_1X_3+  12X_2X_3
     \label{MSM1}
\end{align}
As equation \ref{MSM1} shows the true value of the intercept and the coefficients for $X_1$ and $X_2$ are different from the original model while other coefficients stay the same. The stabilized weight according to equation \ref{sw_eq1} is:
\begin{align}
    IPTW_{SW} = \dfrac{P(X_2) P(X_3)}{P(X_2|X_1,L_2) P(X_3|X_2,L_3)}
\end{align}
Table \ref{MSM_G_Table} compares the $95\%$ CI for the outcome regression and IPTW methods. According to this table, the outcome regression estimate is biased for the coefficients of $X_1$, $X_2$, and the intercept term, while IPTW estimates are unbiased for all the coefficients. This clearly indicates that traditional methods such as outcome regression produce biased estimates in the case of treatment and confounder feedback.
Table \ref{MSM_G_Table2} compares the outcome regression, MSM(IPTW) and non-parametric g-formula. This table represents that both IPTW and g-formula can handle time-varying treatments and confounders feedback.\\
We conclude this section by comparing the IPTW and g-formula: They both can handle time-varying treatments, confounders feedback. They both rely on positivity and conditional exchangeability assumptions (no unmeasured confounders). g-formula is more prone and sensitive to model misspecifications since it requires a correct model for conditional expectation and conditional probability. Non-parametric g-formula becomes very tedious to calculate if the dimension of treatments and confounders increases. Therefore, we believe IPTW is more suitable for educational studies. 

\begin{table}[tbh]
\centering
   \captionsetup{justification=centering}
    \caption{Simulation Results for Time Varying Treatments and Confounders Feedback:\\Comparison of Outcome Regression and MSM in terms of $95\%$ CI}
    \begin{tabular}{|c|c|c|c|}
    \hline
         \textbf{Coefficient} & \textbf{True}  & \textbf{Regression $95\%$ CI}  & \textbf{MSM (IPTW) $95\%$ CI} \\
         \hline
            Intercept & 2.13  & ( 4.92,  5.02 ) & ( 1.910,  3.843 ) \\
            \hline
            $X_1$ &  -2.42   & ( 3.97,  4.11 )  & (-3.351, -0.089 )  \\
            \hline
            $X_2$ &  2.3   &  ( 4.965,  5.120 ) & (-1.0192,  2.3798 )  \\
            \hline
            $X_3$ &   6.0   & ( 5.816,  6.092 ) & ( 4.56,  6.90)  \\
            \hline
            $X_1X_2$ &   10.0   & ( 9.823, 10.047 ) & ( 7.861,  10.604 )  \\
            \hline
            $X_1X_3$ &   11.0    & (10.87, 11.16 )  & ( 7.00, 12.06 ) \\
            \hline
            $X_2X_3$ &   12.0   & (11.78, 12.14 )  & ( 9.56, 13.38 ) \\
            \hline
            $X_1X_2X_3$ &   0   & (-0.1191,  0.2794 ) & (-1.80,  3.04 )  \\
            \hline
    \end{tabular}
    \label{MSM_G_Table}
 \end{table}
\begin{table}[h]
\centering
   \captionsetup{justification=centering}
    \caption{Simulation Results for Time Varying Treatments and Confounders Feedback: \\Comparison of Outcome Regression, MSM and G-formula}
    \begin{tabular}{|c|c|c|c|c|}
    \hline
         \textbf{Coefficient} & \textbf{True} & \textbf{Regression} & \textbf{MSM(IPTW)} & \textbf{G-formula} \\
         \hline
            Intercept & 2.13 & 5.02  &  3.09 & 1.80 \\
            \hline
            $X_1$ &  -2.42   &  4.00  & -1.85 & -2.14  \\
            \hline
            $X_2$ &  2.3   &  4.98   & 1.94 &  1.81\\
            \hline
            $X_3$ &   6.0  &  5.98   & 5.80 &  5.96\\
            \hline
            $X_1X_2$ &   10.0  &  10.04   & 8.65 &  9.84 \\
            \hline
            $X_1X_3$ &   11.0  &  11.04   & 11.22 &  11.01\\
            \hline
            $X_2X_3$ &   12.0  &  12.05  & 12.43 & 12.31 \\
            \hline
            $X_1X_2X_3$ &   0  &  0.09   &0.47 &  0.12 \\
            \hline
    \end{tabular}
    \label{MSM_G_Table2}
 \end{table}
\vspace{2cm}
\section{Conclusion}
In this paper, we proposed various experimental and quasi-experimental designs for educational systems and quantify them using the graphical model and directed acyclic graph (DAG) language. We proposed to model the education system as time-varying treatments, confounders, and time-varying treatments-confounders feedback and derive unbiased causal estimates for the coefficients in fully observed and partially observed confounders. We showed that if we do not control for a sufficient set of confounders, then the causal estimate is biased. We showed that a more intelligent educational system contains feedback between treatments and confounders and provides techniques such as IPTW and g-formula to derive unbiased estimates of the coefficients.\\
Our model can be implemented in a standard college course by having a pretest exam before starting the course to measure students' prerequisite knowledge and including questions about student family, stress, emotional problems, and other factors such as how much time they can spend for the course per week. The pretest score can determine the initial treatment and study plan for students. Then students continue with the intervention and are tested weekly through quizzes to find the best subsequent interventions, such as talking to the TA, instructor, or the counselor depending on the problem they have. We ask the TAs, instructors, and counselors to ask questions during the meeting with students to collect information about the potential confounders and then try to control for these confounders in our analysis. We then have students take an exam (like a midterm or final exam) and follow the methods outlined in this article to evaluate the effectiveness of the joint interventions on the exam score.

\bibliographystyle{unsrt}  
\bibliography{references}  

\end{document}